\documentclass[12pt]{article}%
\usepackage{amsfonts}
\usepackage{amsmath}
\usepackage{geometry}
\usepackage{amssymb}
\usepackage{graphicx}
\usepackage[singlespacing]{setspace}
\usepackage{hyperref}
\usepackage{appendix}%
\providecommand{\U}[1]{\protect\rule{.1in}{.1in}}

\begin{document}

\title{\textbf{Topological Sources of Soliton Mass and Supersymmetry Breaking}}
\author{\textbf{Patrick A. Haas\medskip}\\Department of Physics and Astronomy\\University of Southern California\\Los Angeles, CA 90089, USA\\\medskip\\phaas@usc.edu}
\date{}
\maketitle

\begin{abstract}
\noindent\thispagestyle{empty}We derive the Smarr formulae for two
five-dimensional solutions of supergravity, which are asymptotically $%
\mathbb{R}
^{1,3}\times S^{1}$; in particular, one has a magnetic \textquotedblleft
bolt\textquotedblright\ in its center, and one is a two-center solution. We
show for both spacetimes that supersymmetry -- and so the BPS-bound -- is
broken by the holonomy and how each topological feature of a space-like
hypersurface enters Smarr's mass formula, with emphasis on the ones that give
rise to the stated violation of the BPS-bound. In this light, we question if
any violating extra-mass term in a spacetime with such asymptotics is only
evident in the ADM mass while the Komar mass per s\'{e} \textquotedblleft
tries\textquotedblright\ to preserve BPS. Finally, we derive the cohomological
fluxes for each situation and examine in a more general fashion how the
breaking of supersymmetry -- and so the BPS-bound violation -- is associated
with their topologies. In the second (and more complicated) scenario, we
especially focus on the compact cycle linking the centers, and the
contribution of non-vanishing bulk terms in the mass formula to the breaking
of supersymmetry.\newpage\setcounter{page}{1}

\end{abstract}
\tableofcontents

\section{Introduction}

It has been shown that horizonless solitonic solutions of supergravity can
indeed be constructed purely by means of nontrivial topology. The Smarr
formula has been derived in mutliple works by means of the Komar integral
formalism over cohomology
\cite{Gibbons:2013tqa,Kunduri,Haas:2014,Vercnocke:2015}, one important result
being the role of Chern-Simons terms to only support the topological nature of
the integral.

In this work, we consider two five-dimensional non-BPS solutions of
supergravity which are topologically distinct. The idea in each case is to
compute all contributions from topology and boundary that are flowing into the
total mass formula, to see which pieces precisely cause the breaking of
supersymmetry, by rendering $M\neq\Sigma_{I=1}^{3}Q^{I}$; and which, in
particular, make up $\Delta M=M-\Sigma_{I=1}^{3}Q^{I}$.

In section two, we construct a spacetime with a magnetic \textquotedblleft
bolt\textquotedblright\ in its center and that is asymptotically $%
\mathbb{R}
^{1,3}\times S^{1}$; in the fashion of \cite{Warner:2009,Haas:2015}, we define
a four-dimensional Ricci-flat base space which carries a Euclidean
Schwarzschild metric and magnetic flux from a \textquotedblleft floating
brane\textquotedblright\ ansatz \cite{FloatingBranes2010} for the Maxwell fields.

Topologically, this spacetime can be described by entirely two homological
cycles: The bolt 2-sphere and a non-compact cycle extending from the center to
infinity. In this spirit, the main analysis will be done also in the framework
of intersection homology.

The supersymmetry conditions require that the curvature tensor be either
self-dual or anti-self-dual and tell how this duality has to be correlated
with the one of the magnetic parts of the Maxwell-fields. Since the rotation
group in four-dimensional space decomposes like $SO\left(  4\right)
=SU\left(  2\right)  _{\text{self-dual}}\times SU\left(  2\right)
_{\text{anti-self-dual}}$, only one half of the Killing-spinors would
\textquotedblleft feel\textquotedblright\ space's holonomy and the other half
flat space. In simple examples, this half-flatness, and the preservation or
breaking of supersymmetry can be easily arranged by just changing a sign in
the duality of the fields \cite{GoldsteinKatmadas, BenaWarnerNonBPS,
BenaWarnerMultiCenter}.

However, the curvature tensor for the Euclidean Schwarzschild bolt is neither
self-dual nor anti-self-dual, so one essential BPS-condition is not fulfilled;
but, because the Schwarzschild-geometry is Ricci-flat, the (almost-)BPS
equations of motion are still satisfied \cite{Warner:2009,FloatingBranes2010},
and hence one speaks of an \textquotedblleft almost-BPS\textquotedblright%
-solution. This provides a ground for more general solutions.

Before computing the fluxes and the Komar integral, we will briefly consider a
vastly simplified scenario in which the angular momentum of the running bolt
is set to zero and the five-dimensional warp-factor set to one. The reason for
this is to demonstrate in a very clear and quick way that in a spacetime which
is not asymptotically behaving like $%
\mathbb{R}
^{1,d}$, the theorem $M_{\text{Komar}}=M_{\text{ADM}}$ does in general not
hold anymore. In particular, the Komar mass vanishes under the simplified
conditions, while the ADM mass equals the mass parameter of the Schwarzschild
solution. The latter is responsible for the violation of the BPS-bound
according to $M=Q^{1}+Q^{2}+Q^{3}+m$.

This raises a very important question: Is the extra-mass term violating the
BPS-bound always the difference between the ADM mass and the Komar mass, or,
are there situations in which Komar also reflects terms breaking supersymmetry?

The answer will be illuminated in this work along with the derivation of the
Smarr formula based on the Komar-integral formalism in the sense of
\cite{Gibbons:2013tqa} for the present boundary conditions and for non-BPS
solutions. Further analysis is done within intersection homology -- in
particular, it will be examined how the BPS-bound breaking extra-mass is
composed by the period-integrals of the fields and fluxes in virtue of the
homological cycles. This is to see which intersecting components exist in the
present topology and how they contribute to the breaking of supersymmetry.

We will show that BPS-bound violating mass terms are, in the presence of
compact dimensions, not solely \textquotedblleft seen\textquotedblright\ by
the ADM mass but indeed also arise in the Komar mass from space's topology,
and that this is due to the cohomology dual to the non-compact cycle; in
particular, the latter's intersection with the compact bolt cycle.

This raises the obvious question how supersymmetry-breaking emerges from the
cohomological structure of more complex non-BPS solutions.

In section three, we consider a more general non-supersymmetric solution of
supergravity, this time with two intersecting pieces of compact homology: A
non-extremal center in form of a charged bolt, constructed in a similar
fashion as in section two, and an extremal Gibbons-Hawking center. From
previous works, there are known extremal results for BPS \cite{LuestSabra} and
almost-BPS systems
\cite{GoldsteinKatmadas,BenaWarnerNonBPS,BenaWarnerMultiCenter} -- the
situation considered here is a non-extremal generalization for a non-BPS system.

In particular, we have a running-bolt homology 2-sphere linked with a
Gibbons-Hawking nut by a bubble carrying additional fluxes. This solution was
derived in \cite{BubblingBolt}, first for an arbitrary number of extremal
Gibbons-Hawking centers and then specialized to only one -- so in total two
independent pieces of intersecting compact homology and a non-compact cycle,
much as in section two. The interesting new feature is to look at the
interaction between the bolt and the Gibbons-Hawking nut.\footnote{Another
interesting non-supersymmetric multi-center solution with more than one
homological 2-cycle can be found in \cite{Niehoff:2013mla}}

The main part of the calculations in this section is the explicit examination
of the cohomological fluxes coming from the homological 2-cycles that result
from the bolt and the center-linking bubble. The focus then is on the analysis
of the topological integrals over the harmonics in terms of intersection
homology, in order to see how each cycle contributes to the mass, the charges,
and the BPS-bound violating extra-mass. Although the intersection matrix will
turn out to be rather trivial in section two, a more complicated form is to be
expected in section three. Special emphasis is on the question if and how the
supersymmetry-breaking extra-mass term results in part from space's topology
aside from pure boundary effects.

\section{An almost-BPS spacetime with a magnetic \textquotedblleft running
bolt\textquotedblright}

\subsection{ADM versus Komar}

Before we move on to the computation of the Komar mass, it is essential to
remark its relation to the ADM mass and illustrate this by an easy and quick example.

On a first note, in case of space being asymptotically $%
\mathbb{R}
^{1,d}$, that is, no compactified dimensions, it is known that the two masses
are equal in value. The presence of the $S^{1}$-direction in the spacetime
considered here though, does in fact create a difference, as can be seen by
looking at a strongly simplified version of the current Ricci-flat metric:%
\begin{equation}
ds_{5}^{2}=-dt^{2}+\left(  1-\tfrac{2m}{r}\right)  d\tau^{2}+\left(
1-\tfrac{2m}{r}\right)  ^{-1}dr^{2}+r^{2}\left(  d\theta^{2}+\sin^{2}\theta
d\phi^{2}\right)  . \label{Metric_5D_simplified}%
\end{equation}
This metric is Ricci-flat.

At first, it is important to outline that the ADM mass is in general the more
authoritative measure for the gravitational mass of a system, since Komar, as
stated earlier, requires stationarity.

For a time-like Killing vector, $K=\frac{\partial}{\partial t}$, with dual
1-form $K=g_{00}dt$, the Komar integral becomes%
\begin{equation}
M\sim\int_{X^{3}}\star_{5}dK\sim\int_{X^{3}}\tfrac{\partial}{\partial
r}\left(  g_{00}\right)  \star_{5}\left(  dr\wedge dt\right)  ,
\end{equation}
and from $g_{00}=-1$ we can see directly that the Komar mass vanishes:%
\begin{equation}
M_{\text{Komar}}=0.
\end{equation}

To elaborate on the mass in more detail, we consider orbits in this simplified metric.

The geodesic equations at $\theta=\frac{\pi}{2}$ are:%
\begin{equation}%
\begin{array}
[c]{ccc}%
\tfrac{dt}{d\lambda}=E, & \left(  1-\tfrac{2m}{r}\right)  \frac{d\tau
}{d\lambda}=L_{1}, & r^{2}\frac{d\phi}{d\lambda}=L_{2},
\end{array}
\end{equation}
where $E$, $L_{1}$, and $L_{2}$ are conserved quantities.

The radial equation can be obtained through the normalization condition of the
four-velocity, $u^{\mu}=\frac{dx^{\mu}}{d\lambda}$,%
\begin{equation}
-1=g_{\mu\nu}\tfrac{dx^{\mu}}{d\lambda}\tfrac{dx^{\nu}}{d\lambda}%
=-E^{2}+\tfrac{L_{2}^{2}}{r^{2}}+\left(  1-\tfrac{2m}{r}\right)  ^{-1}\left[
L_{1}^{2}+\left(  \tfrac{dr}{d\lambda}\right)  ^{2}\right]  ,
\end{equation}
and so, keeping only terms of up to first order in $\frac{1}{r}$ at infinity,%
\begin{equation}
\left(  \tfrac{dr}{d\lambda}\right)  ^{2}=\left(  1-\tfrac{2m}{r}\right)
\left(  E^{2}-1\right)  -L_{1}^{2}.
\end{equation}
The radial acceleration at infinity is now:%
\begin{equation}
a^{r}=\tfrac{d^{2}r}{d\lambda^{2}}=\tfrac{E^{2}-1}{\sqrt{\left(  1-\frac
{2m}{r}\right)  \left(  E^{2}-1\right)  -L_{1}^{2}}}\tfrac{m}{r^{2}%
}\rightarrow\tfrac{E^{2}-1}{\sqrt{E^{2}-L_{1}^{2}-1}}\tfrac{m}{r^{2}}.
\end{equation}
Setting off any rotation, $L_{1}=L_{2}=0$, we read off the Keplarian mass
seems to be%
\begin{equation}
M_{\text{Kepler}}=\sqrt{E^{2}-1}\text{ }m,
\end{equation}
which carries a factor of $\sqrt{E^{2}-1}$. In particular, one has $a^{r}=0$
for $E=1$; thus this simple calculation does not directly \textquotedblleft
see\textquotedblright\ the intrinsic mass of the background. To resolve this
issue, we go to a 3+1 description in terms of gravity in 3+1 dimensions to
look at the ADM mass.

Dimensionally reducing the metric along the $\tau$-direction, means to
introduce a conformal scale factor, $\Omega$:%
\begin{equation}
ds_{5}^{2}=\left(  1-\tfrac{2m}{r}\right)  d\tau^{2}+\Omega ds_{4}^{2}.
\end{equation}
The goal is that $ds_{4}^{2}$ will be the metric apparent to observers in 3+1
dimensions. As mentioned above, Komar and ADM mass are equal in value if
asymptotics are $%
\mathbb{R}
^{1,3}$.

The scale factor is necessary to ensure that%
\begin{equation}
\tfrac{1}{G_{5}}\int d^{5}x\sqrt{-g^{\left(  5\right)  }}R^{\left(  5\right)
}=\tfrac{1}{G_{4}}\int d^{4}x\sqrt{-g^{\left(  4\right)  }}R^{\left(
4\right)  }+\text{\textquotedblleft derivatives of scale
factors\textquotedblright.}%
\end{equation}
With these scaling factors one has%
\begin{equation}
g^{\left(  5\right)  }\rightarrow\left(  1-\tfrac{2m}{r}\right)  \Omega
^{4}g^{\left(  4\right)  }\text{ and }R^{\left(  5\right)  }\rightarrow
\Omega^{-1}R^{\left(  4\right)  },
\end{equation}
and hence%
\[
\sqrt{-g^{\left(  5\right)  }}R^{\left(  5\right)  }\rightarrow\sqrt
{1-\tfrac{2m}{r}}\text{ }\Omega\sqrt{-g^{\left(  4\right)  }}R^{\left(
4\right)  }.
\]
Thus one must take%
\begin{equation}
\Omega=\left(  1-\tfrac{2m}{r}\right)  ^{-\frac{1}{2}}.
\end{equation}

Without the scale factor $\Omega$, the four-dimensional Newton constant would
gain radial dependence through multiplication by a power of $\left(
1-\frac{2m}{r}\right)  $.

Rewriting $\left(  \ref{Metric_5D_simplified}\right)  $ in this sense,%
\begin{equation}
ds_{5}^{2}=\left(  1-\tfrac{2m}{r}\right)  d\tau^{2}+\left(  1-\tfrac{2m}%
{r}\right)  ^{-\frac{1}{2}}\left\{  -\left(  1-\tfrac{2m}{r}\right)
^{\frac{1}{2}}dt^{2}+\left(  1-\tfrac{2m}{r}\right)  ^{-\frac{1}{2}}\left[
dr^{2}+\left(  1-\tfrac{2m}{r}\right)  r^{2}\left(  d\theta^{2}+\sin^{2}\theta
d\phi^{2}\right)  \right]  \right\}  ,\nonumber
\end{equation}
leads to the reduced four-dimensional metric:%
\begin{equation}
ds_{4}^{2}=-\left(  1-\tfrac{2m}{r}\right)  ^{\frac{1}{2}}dt^{2}+\left(
1-\tfrac{2m}{r}\right)  ^{-\frac{1}{2}}\left[  dr^{2}+\left(  1-\tfrac{2m}%
{r}\right)  r^{2}\left(  d\theta^{2}+\sin^{2}\theta d\phi^{2}\right)  \right]
,
\end{equation}
and reading off $g_{00}^{\left(  4\right)  }=-\left(  1-\tfrac{2m}{r}\right)
^{\frac{1}{2}}$, yields the stated expression for the ADM-mass:%
\begin{equation}
M_{\text{ADM}}=m.
\end{equation}

This result is the mass parameter of the Schwarzschild solution, which
precisely accounts for the BPS-bound breaking extra-mass term of the solution
within the current simplifications of zero charge.

One concludes that the Komar mass does not detect the breaking of
BPS/super-symmetry while the ADM mass does.

After deriving the Komar mass and Maxwell-charges under the more general
conditions in the following, we will examine the obvious question whether this
is generally true for spacetimes asymptotically behaving like $%
\mathbb{R}
^{1,3}\times S^{1}$, or, possibly $M_{\text{Komar}}=Q^{1}+Q^{1}+Q^{3}+\Delta
M$ for some non-vanishing $\Delta M$.

\subsection{Preliminaries}

In five dimensions, the action is%
\begin{equation}
S=\int\left(  \star_{5}R-Q_{IJ}dX^{I}\wedge\star_{5}dX^{J}-Q_{IJ}F^{I}%
\wedge\star_{5}F^{J}-\tfrac{1}{6}C_{IJK}F^{I}\wedge F^{J}\wedge A^{K}\right)
,
\end{equation}
where $C_{IJK}=\left\vert \epsilon_{IJK}\right\vert $ and $X^{I}$, $I=1,2,3$,
are scalar fields arising from reducing the eleven-dimensional metric,%
\begin{equation}
ds_{11}^{2}=ds_{5}^{2}+\left(  \tfrac{Z_{2}Z_{3}}{Z_{1}^{2}}\right)
^{\frac{1}{3}}\left(  dx_{5}^{2}+dx_{6}^{2}\right)  +\left(  \tfrac{Z_{1}%
Z_{3}}{Z_{2}^{2}}\right)  ^{\frac{1}{3}}\left(  dx_{7}^{2}+dx_{8}^{2}\right)
+\left(  \tfrac{Z_{1}Z_{2}}{Z_{3}^{2}}\right)  ^{\frac{1}{3}}\left(
dx_{9}^{2}+dx_{10}^{2}\right)  , \label{Metric_11D}%
\end{equation}
with the reparametrization,%
\begin{equation}
X^{1}=\left(  \tfrac{Z_{2}Z_{3}}{Z_{1}^{2}}\right)  ^{\frac{1}{3}},\text{
}X^{2}=\left(  \tfrac{Z_{1}Z_{3}}{Z_{2}^{2}}\right)  ^{\frac{1}{3}},\text{
}X^{3}=\left(  \tfrac{Z_{1}Z_{2}}{Z_{3}^{2}}\right)  ^{\frac{1}{3}},
\label{Reparameterization}%
\end{equation}
to fulfill the constraint $X^{1}X^{2}X^{3}=1.$

Moreover, there is a metric for the kinetic terms,%
\begin{equation}
Q_{IJ}=\tfrac{1}{2}\text{diag}\left(  \left(  \tfrac{1}{X^{1}}\right)
^{2},\left(  \tfrac{1}{X^{2}}\right)  ^{2},\left(  \tfrac{1}{X^{3}}\right)
^{2}\right)  , \label{Metric_kinetic terms}%
\end{equation}

We do not allow for cohomology of degree one, so the equations of motion and
symmetry equations correspond to the those in \cite{Haas:2015}, and we only
give a brief summary of the mathematical background emerging the
five-dimensional Komar integral including all the boundary terms.

From varying the action we receive the Einstein and the Maxwell equations
\cite{Gibbons:2013tqa},%
\begin{align}
R_{\mu\nu}  &  =Q_{IJ}\left(  F_{\mu\rho}^{I}F_{\nu}^{J\rho}-\tfrac{1}%
{6}g_{\mu\nu}F_{\rho\sigma}^{I}F^{J\rho\sigma}+\partial_{\mu}X^{I}%
\partial_{\nu}X^{J}\right) \label{Einstein_5D}\\
J_{I\mu}^{CS}  &  =\nabla_{\rho}\left(  Q_{IJ}F_{\text{ \ \ \ }\mu}^{J\rho
}\right)  , \label{Maxwell_5D}%
\end{align}
with the five-dimensional Chern-Simons 1-form current,%
\begin{equation}
J_{I\mu}^{CS}=\tfrac{1}{16}C_{IJK}\bar{\epsilon}_{\mu\rho\sigma\kappa\lambda
}F^{J\rho\sigma}F^{K\kappa\lambda}. \label{CS-current_5D}%
\end{equation}
In differential form language, the Maxwell equations give the known identity
for the dual field strengths,
\begin{equation}
dG_{I}=\tfrac{1}{4}C_{IJK}F^{J}\wedge F^{K}, \label{dG 5D}%
\end{equation}
where%
\begin{equation}
G_{I}=Q_{IJ}\star_{5}F^{J}. \label{G_5D}%
\end{equation}

Eq. $\left(  \ref{Einstein_5D}\right)  $ can be rewritten such that the RHS is
free of any trace terms,%
\begin{equation}
R_{\mu\nu}=Q_{IJ}\left(  \tfrac{2}{3}F_{\mu\rho}^{I}F_{\nu}^{J\rho}%
+\partial_{\mu}X^{I}\partial_{\nu}X^{J}\right)  +\tfrac{1}{6}Q^{IJ}G_{I\mu
\rho\sigma}G_{J\nu}^{\text{ \ \ }\rho\sigma}, \label{Einstein_rearranged_1}%
\end{equation}
especially since this form is much more helpful for the derivation of the
Komar mass formula.

We assume the metric to have a time-like Killing vector, $K^{\mu}$, and can
hence write the five-dimensional mass formula in terms of a Komar integral in
five dimensions,%
\begin{equation}
M=\tfrac{3}{32\pi G_{5}}\int_{X^{3}}\star_{5}dK, \label{Komar integral_5D}%
\end{equation}
where $X^{3}$ is the 3-boundary of the five-dimensional spacetime. Smoothness
of spatial sections, $\Sigma_{4}$, allows in virtue of properties of the
Killing vector to rewrite this formula as an integral over such by $X^{3}$
bound space-like hypersurfaces:%
\begin{equation}
M=\tfrac{3}{32\pi G_{5}}\int_{X^{3}}\star_{5}dK=\tfrac{3}{16\pi G_{5}}%
\int_{\Sigma_{4}}K^{\mu}R_{\mu\nu}d\Sigma^{\nu}.
\end{equation}

Assuming again that the matter fields have the symmetries of the metric, means
them to be invariant under the Lie-derivative along the Killing vector, $K$,%
\begin{equation}
\mathcal{L}_{K}F=0=\mathcal{L}_{K}G, \label{Invariance_5D}%
\end{equation}
from which follow the equations with Cartan's magic formula,%
\begin{equation}
0=d\left(  i_{K}F\right)  \Leftrightarrow i_{K}F^{I}=d\lambda^{I}\text{ and
}i_{K}G_{I}=d\Lambda_{I}-\tfrac{1}{2}C_{IJK}\lambda^{J}F^{K}+H_{I}^{\left(
2\right)  }, \label{F_G_Killing}%
\end{equation}
where $\lambda^{I}$ are magnetostatic potentials of the $G_{I}$ and
electrostatic potentials of the $F^{I}$, respectively; $\Lambda_{I}$ are
globally defined 1-forms and $H_{I}^{\left(  2\right)  }\in H^{2}\left(
\mathcal{M}_{5}\right)  $ closed but not exact 2-forms.

With $\left(  \ref{F_G_Killing}\right)  $ the Einstein equations $\left(
\ref{Einstein_rearranged_1}\right)  $ become%
\begin{equation}
K^{\mu}R_{\mu\nu}=\tfrac{1}{3}\nabla_{\rho}\left(  2Q_{IJ}\lambda^{I}F_{\nu
}^{J\text{ }\rho}+Q^{IJ}\Lambda_{I\sigma}G_{J\nu}^{\text{ \ \ }\rho\sigma
}\right)  +\tfrac{1}{6}Q^{IJ}H_{I}^{\left(  2\right)  \rho\sigma}G_{J\nu
\rho\sigma}. \label{Ricci_Killing_5D}%
\end{equation}

From this follows the Komar mass integral
\cite{Gibbons:1993xt,Sabra:1997yd,Myers:1986un,Peet:2000hn,Gibbons:2013tqa}
over the spatial hypersurface, $\Sigma_{4}$, including the boundary terms over
$X^{3}$:\footnote{The here used convention $\operatorname{div}X=-\delta
X\Rightarrow\delta dZ_{I}=-\hat{\nabla}^{2}Z_{I}$, where $\delta$ is the to
$d$ adjoint exterior derivative, means for here: $\nabla^{2}K_{\mu}=R_{\mu\nu
}K^{\nu}$, so the opposite sign as in \cite{Gibbons:2013tqa}.}%
\begin{equation}
M=-\tfrac{1}{16\pi G_{5}}\left[  \int_{\Sigma_{4}}H_{I}^{\left(  2\right)
}\wedge F^{I}-\int_{X^{3}}\left(  2\lambda^{I}G_{I}-\Lambda_{I}\wedge
F^{I}\right)  \right]  . \label{Komar mass}%
\end{equation}

Since in \cite{Gibbons:2013tqa} the spacetime was assumed to be asymptotic to
$%
\mathbb{R}
^{1,4}$, the boundary integral was taken over $X^{3}=S^{3}$. Throughout this
work, the spacetimes will be asymptotic to $%
\mathbb{R}
^{1,3}\times S^{1}$ and so we have $X^{3}=S_{\infty}^{2}\times S^{1}$.

As we will see later, there is a gauge choice for which $\lambda^{I}$ can be
made zero at infinity. This together with the fact that the $\Lambda_{I}$ are
exact and vanishing at infinity, will prove the Komar mass to be an integral
purely over cohomology, given by the first term in $\left(  \ref{Komar mass}%
\right)  $:%
\begin{equation}
M=-\tfrac{1}{16\pi G_{5}}\int_{\Sigma_{4}}H_{I}^{\left(  2\right)  }\wedge
F^{I}. \label{Komar mass_topology}%
\end{equation}

The Maxwell-charge is computed like:%
\begin{equation}
Q^{I}=-\tfrac{1}{vol\left(  X^{3}\right)  }\int_{X^{3}}G_{I}=-\tfrac{1}%
{32\pi^{2}m}\int_{\Sigma_{4}}dG_{I}=-\tfrac{1}{128\pi^{2}m}C_{IJK}\int
_{\Sigma_{4}}F^{J}\wedge F^{K}. \label{Charge_standard}%
\end{equation}

\subsection{Metric and equations of motion}

The five-dimensional metric, called the \textquotedblleft running
bolt\textquotedblright\ \cite{Warner:2009,Haas:2015} is a time fibration over
Euclidian Schwarzschild:%
\begin{align}
ds_{5}^{2}  &  =-Z^{-2}\left(  dt+k\right)  ^{2}+Zds_{4}^{2}\nonumber\\
&  =-Z^{-2}\left(  dt+k\right)  ^{2}+Z\left[  \left(  1-\tfrac{2m}{r}\right)
d\tau^{2}+\left(  1-\tfrac{2m}{r}\right)  ^{-1}dr^{2}+r^{2}\left(  d\theta
^{2}+\sin^{2}\theta d\phi^{2}\right)  \right]  , \label{Metric_5D_Bolt}%
\end{align}
where $k$ is the angular-momentum 1-form of the running bolt and $Z$ the warp
factor linking the five- and four-dimensional metrics. The coordinate, $\tau$,
results from the Wick-rotation of the time-coordinate in the Euclidian
Schwarzschild base manifold; it parameterizes the $S^{1}$ with periodicity
$\tau\sim\tau+8\pi m$.

The Maxwell fields are set up by the \textquotedblleft floating
brane\textquotedblright\ ansatz \cite{FloatingBranes2010},%
\begin{equation}
A^{I}=-\varepsilon Z_{I}^{-1}\left(  dt+k\right)  +B^{\left(  I\right)  },
\label{A}%
\end{equation}
where $\varepsilon$ is set by the (anti-)self-duality of the fields. The
magnetic field strengths are%
\begin{equation}
\Theta^{\left(  I\right)  }=dB^{\left(  I\right)  }. \label{Theta}%
\end{equation}
The three forms, $Z_{I}$, $\Theta^{\left(  I\right)  }$ and $k$, are
determined through the equations
\cite{GutowskiReall,BenaWarner2005,GoldsteinKatmadas,Warner:2009}:%
\begin{align}
\Theta^{\left(  I\right)  }  &  =\varepsilon\star_{4}\Theta^{\left(  I\right)
},\label{base1}\\
\nabla^{2}Z_{I}  &  =\tfrac{1}{2}\varepsilon C_{IJK}\star_{4}\left[
\Theta^{\left(  J\right)  }\wedge\Theta^{\left(  K\right)  }\right]
,\label{base2}\\
dk+\varepsilon\star_{4}dk  &  =\varepsilon Z_{I}\Theta^{\left(  I\right)  }.
\label{base3}%
\end{align}
Note, that $\left(  \ref{base1}\right)  -\left(  \ref{base3}\right)  $ are
purely represented on the base manifold.

Following the choice of solution for the field strength made in
\cite{Warner:2009},%
\begin{equation}
\Theta^{\left(  I\right)  }=q_{I}\left(  \tfrac{1}{r^{2}}d\tau\wedge
dr+\varepsilon d\Omega_{2}\right)  , \label{Field strength}%
\end{equation}
we have also%
\begin{align}
Z_{I}  &  =1-\tfrac{1}{2m}\tfrac{1}{r}C_{IJK}q_{J}q_{K}\label{Z}\\
k  &  =\mu\left(  r\right)  d\tau=\varepsilon\left(  \tfrac{1}{r}-\tfrac
{1}{2m}\right)  \left[  \Sigma_{I=1}^{3}q_{I}-\tfrac{3}{2m}q_{1}q_{2}%
q_{3}\left(  \tfrac{1}{r}+\tfrac{1}{2m}\right)  \right]  d\tau, \label{k}%
\end{align}
where the $q_{I}$ are $M5$-charges associated with the magnetic field strength
component in $\left(  \ref{Field strength}\right)  $.

It is important to note that exact terms proportional to $d\tau$ have been
chosen such that $k$ vanishes on the bolt, which is essential for regularity
and to remove closed timelike curves. With this choice, the asymptotic limit
of the angular momentum does not vanish but has a finite value:%
\begin{equation}
\mu\overset{r\rightarrow\infty}{\rightarrow}\gamma=-\tfrac{\varepsilon}%
{2m}\left(  \Sigma_{I=1}^{3}q_{I}-\tfrac{3}{4m^{2}}q_{1}q_{2}q_{3}\right)  .
\label{gamma}%
\end{equation}
It is this finite limit which led to the name \textquotedblleft running
bolt\textquotedblright.

Transforming $\left(  \ref{gamma}\right)  $ leads to a formula for the
magnetic charges:%

\begin{equation}
\Sigma_{I=1}^{3}q_{I}=-2\varepsilon m\gamma+\tfrac{3}{4m^{2}}q_{1}q_{2}q_{3}.
\label{M5-charge}%
\end{equation}

\subsection{Topological data}%

\[%
{\parbox[b]{4.657in}{\begin{center}
\includegraphics[
natheight=1.521200in,
natwidth=4.604300in,
height=1.5567in,
width=4.657in
]%
{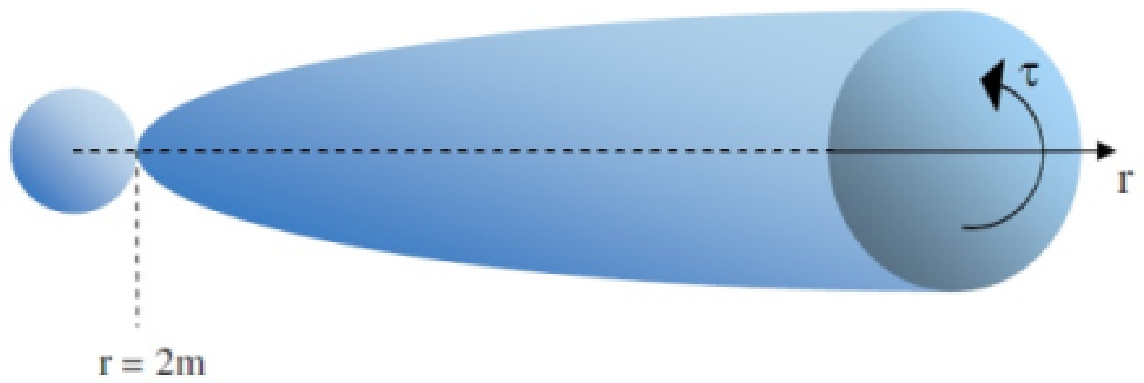}%
\\
Figure 1: Schematic of the homological 2-cycles (from left): bolt and
non-compact cycle
\end{center}}}
\]

The spacetime at hand has entirely two cycles -- the bolt and the non-compact
cycle extending from $r=2m$ to infinity. Each of these carries an independent
cohomological flux: $d\Omega_{2}$, which is carried by the bolt cycle ($B$),
and its dual, $\frac{1}{r^{2}}dr\wedge d\tau$, which is carried by the
non-compact cycle ($\subset$).

The 2-form, $d\Omega_{2}$, is manifestly harmonic. Although its dual can be
written as a total derivative, $\frac{1}{r^{2}}dr\wedge d\tau=d\left(  \left(
\frac{1}{2m}-\frac{1}{r}\right)  d\tau\right)  $, it has a nonvanishing value,
$\frac{1}{2m}d\tau$, at infinity and thus is not exact.

From this, a basis for the cohomology can be directly inferred:%
\begin{align}
v^{B}  &  =\tfrac{1}{4\pi}d\Omega_{2}\label{Basis_bolt}\\
v^{\subset}  &  =\tfrac{1}{4\pi r^{2}}dr\wedge d\tau.
\label{Basis_non-compact cycle}%
\end{align}
It is immediately clear that $v^{B}$ is cohomology. For $v^{\subset}$, on the
other hand, it is not so obvious; but as explained above, its potential,
$\tfrac{1}{4\pi}\left(  \frac{1}{r_{0}}-\tfrac{1}{r}\right)  d\tau$, is either
singular at the bolt or non-vanishing at infinity, depending on $r_{0}$, and
hence cohomological.

From the known results of the fields and the symmetry conditions, we will
derive the cohomological 2-form fluxes. These and the fields will be used to
compute the period-integrals -- the topological \textquotedblleft building
blocks\textquotedblright\ -- to particularly analyse the mass, charges, and
BPS-bound breaking extra-mass within intersection homology.

\subsubsection{Deriving the fluxes}

In the following we derive expressions for the fields and fluxes from the RHS
of $\left(  \ref{Komar mass}\right)  $ to understand their contributations to
the mass formula.

From $\left(  \ref{A}\right)  ,\left(  \ref{Theta}\right)  $ and $\left(
\ref{Field strength}\right)  $ follows the Maxwell-field strength,
$F^{I}=dA^{I}$, which decomposes into an exact and a (on the base manifold)
harmonic part,%
\begin{equation}
F^{I}=d\hat{A}^{I}+\varepsilon q_{I}d\Omega_{2}-\left(  2m\varepsilon
\gamma+q_{I}\right)  \tfrac{1}{r^{2}}dr\wedge d\tau, \label{F_decomposed}%
\end{equation}
where%
\begin{equation}
\hat{A}^{I}=\varepsilon\left[  -Z_{I}^{-1}\mu+\gamma\left(  1-\tfrac{2m}%
{r}\right)  \right]  d\tau+\text{\textquotedblleft}dt\text{\textquotedblright%
}. \label{A2}%
\end{equation}
Since the time-coordinate, $t$, is not part of the base space, its 1-form,
$dt$, is irrelevant for the cohomology.

Note that the $\hat{A}^{I}$ vanish at the bolt and at infinity and are thus
globally smooth.

Choosing the Killing vector like%
\begin{equation}
K=\tfrac{\partial}{\partial t},
\end{equation}
we get from $\left(  \ref{F_G_Killing}a\right)  $:%
\begin{equation}
\lambda^{I}=\varepsilon Z_{I}^{-1}-\beta^{I}, \label{lambda_small_general}%
\end{equation}
where $\beta^{I}$ are constants.

It is directly obvious that the choice%
\begin{equation}
\beta^{I}=\varepsilon, \label{gauge-choice}%
\end{equation}
causes the $\lambda^{I}$ to vanish at infinity, where $Z_{I}\rightarrow1$.
This way the term $2\lambda^{I}G_{I}$ drops from the boundary integral of
$\left(  \ref{Komar mass}\right)  $. Also, the below computed exact 1-forms,
$\Lambda_{I}$, from $\left(  \ref{Ricci_Killing_5D}\right)  $ have to fall off
at infinity; and so the Komar mass is rendered purely topological.

With $\left(  \ref{A}\right)  $ and $\left(  \ref{G_5D}\right)  $ we have%
\begin{align}
G_{I}  &  =\tfrac{1}{2}\left[  -\varepsilon r^{2}\left(  1-\tfrac{2m}%
{r}\right)  Z_{I}^{\prime}+\varepsilon r^{2}Z_{I}Z^{-3}\mu\mu^{\prime}%
+q_{I}Z_{I}^{2}Z^{-3}\mu\right]  d\tau\wedge d\Omega_{2}\nonumber\\
&  +\tfrac{1}{2}Z_{I}Z^{-3}\left(  \varepsilon r^{2}\mu^{\prime}+q_{I}%
Z_{I}\right)  dt\wedge d\Omega_{2}+\tfrac{\varepsilon}{2r^{2}}q_{I}Z_{I}%
^{2}Z^{-3}dt\wedge d\tau\wedge dr, \label{G 5D}%
\end{align}
and find from this together with $\left(  \ref{lambda_small_general}\right)
$,%
\begin{equation}
i_{K}G_{I}+\tfrac{1}{2}C_{IJK}\lambda^{J}F^{K}=-\tfrac{1}{2}C_{IJK}\beta
^{J}F^{K}-\tfrac{1}{2}d\left(  Z_{I}Z^{-3}\left(  dt+\mu d\tau\right)
\right)  , \label{Fluxes_combined}%
\end{equation}
which is a manifestly closed expression.

Using $\left(  \ref{F_G_Killing}b\right)  $, we put all exact pieces of
$\left(  \ref{Fluxes_combined}\right)  $ into $d\Lambda_{I}$ so that we have:%
\begin{equation}
\Lambda_{I}=-\tfrac{1}{2}\left[  Z_{I}Z^{-3}\mu-\gamma\left(  1-\tfrac{2m}%
{r}\right)  -\varepsilon C_{IJK}\beta^{J}\left(  Z_{K}^{-1}\mu-\gamma\left(
1-\tfrac{2m}{r}\right)  \right)  \right]  d\tau+\text{\textquotedblleft%
}dt\text{\textquotedblright}, \label{lambda_big_general}%
\end{equation}
which is smooth at the bolt and vanishes at infinity.

The remaining terms in $\left(  \ref{Fluxes_combined}\right)  $ sum up to the
2-form harmonic:%
\begin{equation}
H_{I}^{\left(  2\right)  }=-\tfrac{\varepsilon}{2}C_{IJK}q_{J}\beta^{K}%
d\Omega_{2}-\left[  m\gamma-\tfrac{1}{2}C_{IJK}\left(  q_{J}+2m\varepsilon
\gamma\right)  \beta^{K}\right]  \tfrac{1}{r^{2}}dr\wedge d\tau.
\label{H2_general}%
\end{equation}

As mentioned earlier, the 2-form, $\frac{1}{r^{2}}dr\wedge d\tau$, has
nonvanishing potential at infinity and thus is cohomological.

\subsubsection{Intersection homology}

The intersection technique relates an integral of two wedged 2-forms over the
whole four-dimensional base space to integrals of the single 2-forms over the
homological 2-cycles.

Applying the gauge choice $\left(  \ref{gauge-choice}\right)  $, $\beta
^{I}=\varepsilon$, rendering the Komar mass purely topological, the
period-integrals, forming the topological \textquotedblleft building
blocks\textquotedblright,\ ammount to:%
\begin{align*}
&
\begin{tabular}
[c]{|l|l|l|}\hline
& $C_{B}=S_{r=2m}^{2}$ & $C_{\subset}=S^{1}\times\left[  2m,\infty\right[
$\\\hline
$\int F^{I}$ & $4\pi\varepsilon q_{I}$ & $-4\pi\left(  2m\varepsilon
\gamma+q_{I}\right)  $\\\hline
$\int H_{I}^{\left(  2\right)  }$ & $-2\pi\left(  \Sigma_{J=1}^{3}q_{J}%
-q_{I}\right)  $ & $2\pi\left[  \varepsilon\left(  \Sigma_{J=1}^{3}q_{J}%
-q_{I}\right)  +2m\gamma\right]  $\\\hline
\end{tabular}
\\
&  \text{Table 1}\colon\text{ Integrals of the 2-forms over the 2-cycles}%
\end{align*}

It is instructive to introduce a canonical integer basis for the cohomology,%
\begin{equation}
\int_{C_{A}}v^{A^{\prime}}=\delta_{A}^{A^{\prime}}\text{ and }\int_{\Sigma
_{4}}v^{A}\wedge v^{A^{\prime}}=I^{AA^{\prime}},
\end{equation}
with%
\begin{equation}
F^{I}=\sigma_{A}^{I}v^{A}\text{ and }H_{I}^{\left(  2\right)  }=\tilde{\sigma
}_{I,A}v^{A},
\end{equation}
where $\sigma_{A}^{I}$ and $\tilde{\sigma}_{I,A}$ ($A=B,\subset$) are
precisely the entries of the above tables (the \textquotedblleft building
blocks\textquotedblright), and $I^{AA^{\prime}}=I^{A^{\prime}A}$ is the
inverse intersection matrix.

The choice of the cohomological basis $\left(  \ref{Basis_bolt}\right)
-\left(  \ref{Basis_non-compact cycle}\right)  $ manifestly fulfills the
above-stated orthonormality condition.

The integrals for the mass and charge formulae become then%
\begin{align}
\int_{\Sigma_{4}}H_{I}^{\left(  2\right)  }\wedge F^{I}  &  =16\pi^{2}%
\Sigma_{I=1}^{3}\left(  C_{IJK}q_{J}q_{K}+3\varepsilon m\gamma q_{I}\right)
=\tilde{\sigma}_{I,A}\sigma_{A^{\prime}}^{I}I^{AA^{\prime}}\label{H_F}\\
C_{IJK}\int_{\Sigma_{4}}F^{J}\wedge F^{K}  &  =-32\pi^{2}\varepsilon
C_{IJK}q_{J}\left(  2m\varepsilon\gamma+q_{K}\right)  =C_{IJK}\sigma_{A}%
^{J}\sigma_{A^{\prime}}^{K}I^{AA^{\prime}}. \label{F_F}%
\end{align}
In particular, this means that the integrals need to be reproduced by
composing the products of the period-integrals, $\sigma_{A}^{I}$ and
$\tilde{\sigma}_{I,A}$, by the integer coefficients of $I^{AA^{\prime}}$ in
the sense of the last expressions, which can only be achieved by%
\begin{equation}
I^{AA^{\prime}}=I_{AA^{\prime}}=\left(
\begin{array}
[c]{cc}%
0 & 1\\
1 & 0
\end{array}
\right)  .
\end{equation}
This is merely a trivial one-time intersection between the bolt and the
non-compact cycle, proving that there is no self-intersecting homology at hand.

Note: If we take the building blocks from table 1 and compute $\sigma_{A}%
^{I}v^{A}$, then we get%
\begin{equation}
F_{\text{harmonic}}^{I}=\varepsilon q_{I}d\Omega_{2}-\left(  2m\varepsilon
\gamma+q_{I}\right)  \tfrac{1}{r^{2}}dr\wedge d\tau,
\end{equation}
but it is%
\begin{equation}
F^{I}-F_{\text{harmonic}}^{I}=d\hat{A},
\end{equation}
what is cohomologous to the total Maxwell-field strength, where $\hat{A}$ is a
global 1-form falling off at infinity.

In this spirit, it is now easy to compute the total Komar mass and Maxwell-charge.

\subsection{Mass, charges, and the breaking of supersymmetry}

In this section, we will evaluate the expressions for the Komar mass and the
Maxwell-charges by means of the just introduced intersection homology method.

The fact that $\Lambda_{I}\rightarrow0$ at infinity and the gauge choice,%
\begin{equation}
\beta^{I}=\varepsilon,
\end{equation}
for which the function $\lambda^{I}$ goes zero according to $\left(
\ref{lambda_small_general}\right)  $, make the boundary integral drop out of
$\left(  \ref{Komar mass}\right)  $, rendering the Komar mass an integral
purely over cohomology $\left(  \ref{Komar mass_topology}\right)  $:%
\begin{equation}
M=-\tfrac{1}{16\pi G_{5}}\int_{\Sigma_{4}}H_{I}^{\left(  2\right)  }\wedge
F^{I}. \label{Komar mass_reduced}%
\end{equation}

From $\left(  \ref{Komar mass_reduced}\right)  $ and $\left(
\ref{Charge_standard}\right)  $ we get with $\left(  \ref{H_F}\right)
-\left(  \ref{F_F}\right)  $ the mass and charges:%
\begin{align}
M  &  =-\tfrac{\pi}{G_{5}}\Sigma_{I=1}^{3}\left(  3m\varepsilon\gamma
q_{I}+C_{IJK}q_{J}q_{K}\right) \label{M}\\
Q^{I}  &  =\tfrac{\varepsilon}{4m}C_{IJK}q_{J}\left(  2m\varepsilon
\gamma+q_{K}\right)  . \label{Q_summed}%
\end{align}

From that follows%
\begin{equation}
M=-\tfrac{\varepsilon\pi m}{G_{5}}\left(  4\Sigma_{I=1}^{3}Q^{I}-\gamma
\Sigma_{I=1}^{3}q_{I}\right)  ,
\end{equation}
and hence the BPS-bound violating extra-mass%
\begin{align}
\Delta M  &  =M+\tfrac{4\varepsilon\pi m}{G_{5}}\Sigma_{I=1}^{3}Q^{I}%
=\tfrac{\varepsilon\pi m\gamma}{G_{5}}\Sigma_{I=1}^{3}q_{I}%
\label{Extra-mass_running-bolt}\\
&  =-\tfrac{1}{16\pi G_{5}}\left(  \int_{\Sigma_{4}}H_{I}^{\left(  2\right)
}\wedge F^{I}+\tfrac{\varepsilon}{2}\Sigma_{I=1}^{3}C_{IJK}\int_{\Sigma_{4}%
}F^{J}\wedge F^{K}\right) \\
&  =-\tfrac{1}{16\pi G_{5}}\sigma_{A^{\prime}}^{K}\left(  \tilde{\sigma}%
_{K,A}+\tfrac{\varepsilon}{2}\Sigma_{I=1}^{3}C_{IJK}\sigma_{A}^{J}\right)
I^{AA^{\prime}}.
\end{align}
Interestingly, the breaking of supersymmetry is caused by the total
$M5$-charges, $q_{I}$.

Very important note: The factored term,%
\begin{equation}
\chi_{K,A}=\tilde{\sigma}_{K,A}+\tfrac{\varepsilon}{2}\Sigma_{I=1}^{3}%
C_{IJK}\sigma_{A}^{J},
\end{equation}
considered at each cycle separately,%
\begin{align}
\chi_{K,B}  &  =0\\
\chi_{K,\subset}  &  =-4\pi m\gamma,
\end{align}
vanishes identically for the bolt-cycle, $A=B$, so all contribution to the
breaking of supersymmetry comes from the non-compact cycle, $A=\subset$:%
\begin{align}
\Delta M  &  =-\tfrac{1}{16\pi G_{5}}\sigma_{B}^{K}\left(  \tilde{\sigma
}_{K,\subset}+\tfrac{\varepsilon}{2}\Sigma_{I=1}^{3}C_{IJK}\sigma_{\subset
}^{J}\right)  I^{B\subset}\\
&  =-\tfrac{1}{16\pi G_{5}}\Sigma_{K=1}^{3}4\pi\varepsilon q_{K}\left(  -4\pi
m\gamma\right) \\
&  =\tfrac{\varepsilon\pi m\gamma}{G_{5}}\Sigma_{K=1}^{3}q_{K}.
\end{align}

\section{Topological contributions to the BPS-bound violation in a 2-center
solution}

\subsection{The 2-center solution}%

\[%
{\parbox[b]{4.9701in}{\begin{center}
\includegraphics[
natheight=1.249600in,
natwidth=4.916500in,
height=1.2851in,
width=4.9701in
]%
{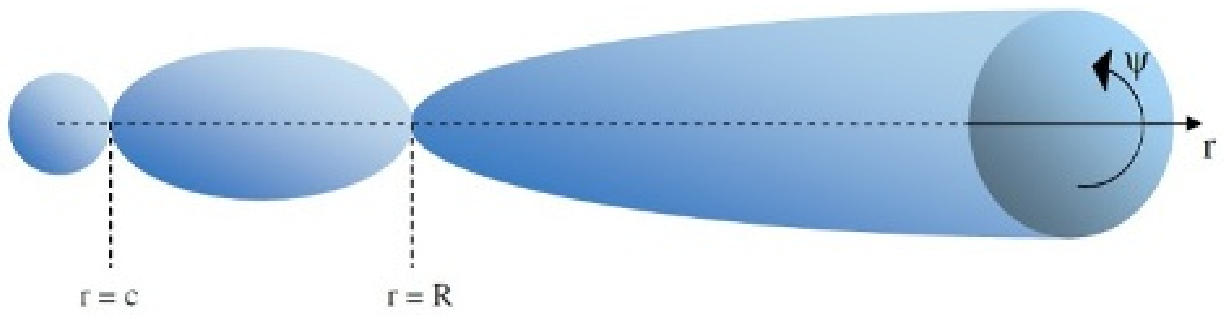}%
\\
Figure 2: Schematic of the homological 2-cycles (from left): bolt, bubble, and
non-compact cycle
\end{center}}}
\]

In the following we outline the main results of the non-supersymmetric
2-center solution given in \cite{BubblingBolt}.

The geometry of the spacetime and the three $U\left(  1\right)  $-gauge fields
are a solution of the Einstein-Maxwell equations in the floating brane ansatz
\cite{FloatingBranes2010} within five-dimensional ungauged supergravity. The
stationary spacetime carries a Killing vector, $K=\frac{\partial}{\partial t}%
$, and has the metric:%
\begin{equation}
ds_{5}^{2}=-\left(  \tfrac{1}{2}LL_{a}L^{a}\right)  ^{-\frac{2}{3}}\left(
dt+\hat{k}\right)  ^{2}+\left(  \tfrac{1}{2}LL_{a}L^{a}\right)  ^{\frac{1}{3}%
}ds_{4}^{2}, \label{Metric_5D}%
\end{equation}
where the functions $L$ and $L_{a}$ and the angular momentum 1-form $\hat{k}$
will be defined below.

The parameter $a$ in $\left(  \ref{Metric_5D}\right)  $ counts vector
multiplets; like in \cite{Warner:2009,Gibbons:2013tqa,Haas:2015} we choose to
have two ($a=2,3$) to have a total of three Maxwell-charges, $A^{I}$
($I=1,2,3$). This enables non-trivial Chern-Simons interactions.

However, like \cite{BubblingBolt} we use the STU truncation in which the
fields and fluxes with index $I=1$ are treated in a different way than $I=a$.
In detail, the raise of the latter index, $K_{a}=\eta_{ab}K^{b}$, happens with
an $SO\left(  1,1\right)  $ metric following from the non-zero Chern-Simons
coupling $C_{1ab}=\eta_{ab}=\left(
\begin{array}
[c]{cc}%
0 & 1\\
1 & 0
\end{array}
\right)  $.

As for the running bolt solution from the last section and Gibbons-Hawking
metrics, the four-dimensional base space is a $U\left(  1\right)  $-fibration
over a 3-manifold which is asymptotically $%
\mathbb{R}
^{3}$ at infinity, rendering the asymptotics of the whole spatial base
$S^{1}\times%
\mathbb{R}
^{3}$ at infinity.

In particular, the 3-manifold is parameterized by $\left(  r,\theta
,\phi\right)  $ and the fiber by $\psi$. As in the last section, the latter
defines a compact spatial dimension -- the $S^{1}$ --, this time with
periodicity $\psi\sim\psi+4\pi k$, where $k$ is the scale parameter of the
$S^{1}$.

The special topology considered here, is constituted by a non-extremal charged
bolt at $r=c$ and an extremal Gibbons-Hawking center at $r=R>c$ and $\theta
=0$. This leads to a non-extremal and non-BPS generalization of known extremal
results for BPS \cite{LuestSabra} and almost-BPS systems
\cite{GoldsteinKatmadas,BenaWarnerNonBPS,BenaWarnerMultiCenter}.

The most crucial aspect of this topology is that, the $\psi$-fiber pinches off
at two locations -- the bolt and the Gibbons-Hawking center. This $\psi$-fiber
along an interval between $r=c$ and $r=R$ at $\theta=0$ defines the new
non-trivial compact homology cycle.

The metric of the four-dimensional base manifold is%
\begin{equation}
ds_{4}^{2}=V^{-1}\left(  d\psi+\omega^{0}\right)  ^{2}+V\left[  dr^{2}+\left(
r^{2}-c^{2}\right)  \left(  d\theta^{2}+\sin^{2}\theta d\phi^{2}\right)
\right]  , \label{Metric_4D}%
\end{equation}
where $V$ has poles at $r=c$ and $\left(  r,\theta\right)  =\left(
R,0\right)  $. This causes the $\psi$-fiber to pinch off, but leaves the
metric smooth.

This leads to the main difference to the running-bolt spacetime of the last
section, which had only one pinch-off point for the circle-fibration.

As we will see in detail, the fact that $\psi$ pinches off at \textit{two}
centers generates a new homological cycle defined by the fiber and the radial
interval between the two pinch-off points.

In particular, if a periodic coordinate pinches off at only one existing
center, like in the running-bolt\ solution, it can be fixed at the bolt at the
cost of creating flux that does not vanish at infinity and hence giving rise
to a non-compact cycle. Two pinch-off points, as considered here, give rise to
both a further compact as well as the non-compact cycle.

Entirely, there are three independent homological 2-cycles (fig. 1): The
bolt-sphere at $r=c$; the bubble-cycle being the $\psi$-fibered radial line
along the positive $z$-axis, connecting the bolt's north pole and the
Gibbons-Hawking center at $\left(  r,\theta\right)  =\left(  R,0\right)  $;
the non-compact cycle, also running radially along the positive $z$-axis from
the Gibbons-Hawking center, $z=R$, towards infinity, fibred by the $\psi$-circles.

The appendix gives more information on the functions used here\footnote{The
functions used here are taken from eqs. $\left(  2.25\right)  $, $\left(
2.26\right)  $, $\left(  3.55\right)  $, $\left(  4.2\right)  -\left(
4.6\right)  $ of \cite{BubblingBolt}.} along with their asymptotics -- they
are very complicated, which reinforces the interest in cohomology --, but we
will need some details here\footnote{In \cite{BubblingBolt} the functions were
equipped with an extra-parameter, $n_{A}$, that can be set equal to $1$, which
we do throughout this section.}:%
\begin{align}
V\left(  r,\theta\right)   &  =\tfrac{r+m_{-}}{2\left(  r^{2}-c^{2}\right)
}\left(  r+m_{+}-\tfrac{2k}{R+m_{-}}\tfrac{Rr-c^{2}\cos\theta}{\sqrt{r_{1}%
^{2}-c^{2}\sin^{2}\theta}}\right) \\
\hat{k}  &  =\omega-\tfrac{M}{V}\left(  d\psi+\omega^{0}\right) \\
\omega^{0}\left(  r,\theta\right)   &  =-\tfrac{1}{2}\left[  \left(
m_{+}-m_{-}\right)  \cos\theta+\tfrac{2k}{R+m_{-}}\tfrac{R^{2}-m_{-}r-R\left(
r-m_{-}\right)  \cos\theta-c^{2}\sin^{2}\theta}{\sqrt{r_{1}^{2}-c^{2}\sin
^{2}\theta}}\right]  d\phi\\
\omega\left(  r,\theta\right)   &  =-\tfrac{e_{-}R}{2\left(  R+m_{-}\right)
^{2}}u_{a}u^{a}\left[  \left(  1-\tfrac{r+R}{r_{1}}\right)  \left(
1-\cos\theta\right)  +\tfrac{c^{2}}{Rr_{1}}\sin^{2}\theta\right]  d\phi\\
L_{a}\left(  r,\theta\right)   &  =\tfrac{\left(  r+m_{-}\right)  \left(
c^{2}+m_{-}r\right)  }{2m_{-}\left(  r^{2}-c^{2}\right)  }\tfrac{l_{a}}%
{V}+u_{a}\\
L\left(  r,\theta\right)   &  =\tfrac{e_{-}^{2}}{2m_{-}^{2}}\tfrac{1}{V}%
l_{a}l^{a}-\tfrac{e_{-}^{2}}{c^{2}\left(  c+m_{-}\right)  ^{2}}\tfrac
{f_{1}r+f_{2}}{\left(  m_{-}+r\right)  \left(  m_{-}+R\right)  }u_{a}u^{a}\\
M\left(  r,\theta\right)   &  =-\tfrac{e_{-}}{2m_{-}}l_{a}L^{a}+\tfrac{e_{-}%
}{2\left(  m_{-}+R\right)  }\left[  \tfrac{R-r}{m_{-}+r}V+\tfrac{\left(
c^{2}+m_{-}r\right)  \left(  f_{1}r+f_{2}\right)  }{2c^{2}\left(
c+m_{-}\right)  ^{2}\left(  r^{2}-c^{2}\right)  }\right]  u_{a}u^{a},
\end{align}
where $r_{1}=\sqrt{r^{2}+R^{2}-2Rr\cos\theta}$ is the distance measured from
the Gibbons-Hawking center; $m_{-}$, $m_{+}$, $e_{-}$, $l_{a}$, $u_{a}$,
$f_{1}$, and $f_{2}$ are parameters of the solution which are non-trivially
interrelated (see appendix A).

The function $V$ has two singularities -- one at the bolt, going like
$\frac{k}{r-c}$, and one at the Gibbons-Hawking point, going like $-\tfrac
{k}{\left\vert r-R\right\vert }$ at $\theta=0$; in the metric $\left(
\ref{Metric_4D}\right)  $ it poses a coordinate singularity and so does not
harm regularity at the centers.\footnote{For a much more detailed outline of
the regularity analysis, see \cite{BubblingBolt}.}

Like the geometry, the Maxwell-fields are solutions to the Einstein-Maxwell
equations; their potentials in the floating brane ansatz are:%
\begin{align}
A^{1}  &  =\tfrac{1}{L}\left(  dt+\hat{k}\right)  -\tfrac{1}{2e_{-}}\left[
\tfrac{\left(  r+m_{-}\right)  \left(  c^{2}+m_{-}r\right)  }{V\left(
r^{2}-c^{2}\right)  }\left(  d\psi+\omega^{0}\right)  +\left(  c^{2}-m_{-}%
^{2}\right)  \cos\theta d\phi\right] \label{A^1}\\
A^{a}  &  =\tfrac{1}{L_{a}}\left(  dt+\hat{k}\right)  -\tfrac{e_{-}}{m_{-}%
}\tfrac{l^{a}}{V}\left(  d\psi+\omega^{0}\right)  +e_{-}u^{a}\left[  \tfrac
{2}{r+m_{-}}\left(  d\psi+\omega^{0}\right)  -\left(  \cos\theta+\tfrac
{2k}{R+m_{-}}\tfrac{r-R\cos\theta}{\sqrt{r_{1}^{2}-c^{2}\sin^{2}\theta}%
}\right)  d\phi\right]  , \label{A^a}%
\end{align}
where in both cases the first part represents the terms which are globally
defined -- yet not exact, for they do not fall off at infinity.

From this follow their field-strengths, $F^{I}=dA^{I}$,%
\begin{align}
F^{1}  &  =d\left(  \tfrac{1}{L}\left(  dt+\hat{k}\right)  \right)
+Z_{1}\left[  V\left(  r^{2}-c^{2}\right)  d\Omega_{2}-dr\wedge\left(
d\psi+\omega^{0}\right)  \right]  +Z_{2}\left[  d\theta\wedge\left(
d\psi+\omega^{0}\right)  +V\sin\theta dr\wedge d\phi\right] \label{F^1}\\
F^{a}  &  =d\left(  \tfrac{1}{L_{a}}\left(  dt+\hat{k}\right)  -\tfrac{e_{-}%
}{m_{-}}\tfrac{l^{a}}{V}\left(  d\psi+\omega^{0}\right)  \right)
+\tfrac{2e_{-}u^{a}}{\left(  r+m_{-}\right)  ^{2}}\left[  V\left(  r^{2}%
-c^{2}\right)  d\Omega_{2}-dr\wedge\left(  d\psi+\omega^{0}\right)  \right]  ,
\label{F^a}%
\end{align}
where the terms have been sorted according to the ones of the potentials.

Extracting the topological bits out of the globally defined terms, to make
them exact, can be done with help of the cohomological basis which we derive
later. Adding them to the other terms, however, would render them not harmonic
anymore, so the present topology has both self-dual and anti-self-dual parts.

The functions $Z_{1,2}$, have the form:%
\begin{align}
Z_{1}  &  =\tfrac{\left(  c^{2}+m_{-}r\right)  \left(  r+m_{-}\right)
}{2e_{-}V\left(  r^{2}-c^{2}\right)  }\left[  \tfrac{r+m_{-}}{2V\left(
r^{2}-c^{2}\right)  }-\tfrac{m_{-}}{c^{2}+m_{-}r}+\left(  1-\tfrac{\left(
r+m_{-}\right)  \left(  r+m_{+}\right)  }{2V\left(  r^{2}-c^{2}\right)
}\right)  \left(  \tfrac{R}{Rr-c^{2}\cos\theta}+\tfrac{R\cos\theta-r}%
{r_{1}^{2}-c^{2}\sin^{2}\theta}\right)  \right] \label{Z1}\\
Z_{2}  &  =-\tfrac{\left(  c^{2}+m_{-}r\right)  \left(  r+m_{-}\right)
}{2e_{-}V}\left(  1-\tfrac{\left(  r+m_{-}\right)  \left(  r+m_{+}\right)
}{2V\left(  r^{2}-c^{2}\right)  }\right)  \left(  \tfrac{R}{Rr-c^{2}\cos
\theta}+\tfrac{R\cos\theta-r}{r_{1}^{2}-c^{2}\sin^{2}\theta}\right)
\tfrac{\sin\theta}{R-r\cos\theta}. \label{Z2}%
\end{align}

The dual field strengths have to be computed by%
\begin{align}
G_{1}  &  =\tfrac{1}{2}L^{\frac{4}{3}}\left(  \tfrac{1}{2}L_{c}L^{c}\right)
^{-\frac{2}{3}}\star_{5}F^{1}\label{G_1}\\
G_{a}  &  =\tfrac{1}{2}L^{-\frac{2}{3}}\left(  \tfrac{1}{2}L_{c}L^{c}\right)
^{\frac{4}{3}}\left(  L^{a}\right)  ^{-2}\star_{5}F^{a}, \label{G_a}%
\end{align}
where the prefactors correspond to the $Q_{IJ}$ from $\left(  \ref{G_5D}%
\right)  $, composed in the same manner as in $\left(
\ref{Metric_kinetic terms}\right)  $ of the scalars of the solution:%
\[
X^{1}=L^{-\frac{2}{3}}\left(  \tfrac{1}{2}L_{a}L^{a}\right)  ^{\frac{1}{3}%
}\text{ and }X^{a}=L^{\frac{1}{3}}\left(  \tfrac{1}{2}L_{b}L^{b}\right)
^{-\frac{2}{3}}L^{a}.
\]

The derivation of the Komar mass formula was done in the last section in
$\left(  \ref{Komar mass}\right)  $; the three $U\left(  1\right)  $-charges
relevant for this section, are given by the properly normalized formula:%
\begin{equation}
Q^{I}=-\tfrac{1}{vol\left(  X^{3}\right)  }\int_{X^{3}}G_{I}=-\tfrac{1}%
{16\pi^{2}k}\int_{\Sigma_{4}}dG_{I}=-\tfrac{1}{64\pi^{2}k}C_{IJK}\int
_{\Sigma_{4}}F^{J}\wedge F^{K}, \label{Charge-formula}%
\end{equation}
where the last step follows from the equations of motion for the
Maxwell-fields $\left(  \ref{dG 5D}\right)  $.

\subsection{Deriving the fluxes}

In the following, we derive expressions for the fields and fluxes to
understand their contributions to the mass formula. As above, we write the
timelike Killing vector near infinity like $K=\tfrac{\partial}{\partial t}$.

From $\left(  \ref{F^1}\right)  -\left(  \ref{F^a}\right)  $ we get%
\begin{align}
i_{K}F^{1}  &  =-d\left(  \tfrac{1}{L}\right) \\
i_{K}F^{a}  &  =-d\left(  \tfrac{1}{L_{a}}\right)  ,
\end{align}
\bigskip and so with $\left(  \ref{F_G_Killing}a\right)  $:%
\begin{align}
\lambda^{1}  &  =\beta^{1}-\tfrac{1}{L}\label{lambda_small^1}\\
\lambda^{a}  &  =\beta^{a}-\tfrac{1}{L_{a}}, \label{lambda_small^a}%
\end{align}
where the $\beta^{I}$ are freely-choosable constants. We will later fix them
though, so that $\lambda^{I}\rightarrow0$ at infinity.

From $\left(  \ref{F^1}\right)  -\left(  \ref{G_a}\right)  $ and $\left(
\ref{lambda_small^1}\right)  -\left(  \ref{lambda_small^a}\right)  $ we find,
decomposing $\left(  \ref{F_G_Killing}b\right)  $ according to the STU
truncation:%
\begin{align}
\Gamma_{1}  &  =\tfrac{e_{-}}{\left(  r+m_{-}\right)  ^{2}}\left(  u_{a}%
\beta^{a}-2\right)  \left[  V\left(  r^{2}-c^{2}\right)  d\Omega_{2}%
-dr\wedge\left(  d\psi+\omega^{0}\right)  \right] \nonumber\\
&  +d\left[  \tfrac{1}{L_{c}L^{c}}\left(  \beta^{a}L_{a}-1\right)  \left(
dt+\hat{k}\right)  -\tfrac{e_{-}}{2m_{-}V}l_{a}\beta^{a}\left(  d\psi
+\omega^{0}\right)  \right] \label{Fluxes1}\\
\Gamma_{a}  &  =\left(  \tfrac{e_{-}}{\left(  r+m_{-}\right)  ^{2}}u_{a}%
\beta^{1}+\tfrac{1}{2}\eta_{ab}\beta^{b}Z_{1}\right)  \left[  V\left(
r^{2}-c^{2}\right)  d\Omega_{2}-dr\wedge\left(  d\psi+\omega^{0}\right)
\right] \nonumber\\
&  +\tfrac{1}{2}\eta_{ab}\beta^{b}Z_{2}\left[  V\sin\theta dr\wedge
d\phi+d\theta\wedge\left(  d\psi+\omega^{0}\right)  \right] \label{Fluxes2}\\
&  +d\left[  \tfrac{1}{2}\left(  \tfrac{1}{L^{a}}\beta^{1}+\tfrac{1}{L}%
\eta_{ab}\beta^{b}-\tfrac{1}{LL^{a}}\right)  \left(  dt+\hat{k}\right)
-\tfrac{e_{-}}{2m_{-}V}l_{a}\beta^{1}\left(  d\psi+\omega^{0}\right)  \right]
,\nonumber
\end{align}
where we wrote for short,%
\begin{align}
\Gamma_{1}  &  =i_{K}G_{1}+\tfrac{1}{2}\eta_{ab}\lambda^{a}F^{b}\\
\Gamma_{a}  &  =i_{K}G_{a}+\tfrac{1}{2}\eta_{ab}\left(  \lambda^{1}%
F^{b}+\lambda^{b}F^{1}\right)  .
\end{align}
These formulae suggest obvious choices for the analogues of $\left(
\ref{lambda_big_general}\right)  $ and $\left(  \ref{H2_general}\right)  $:%
\begin{align}
\tilde{H}_{1}^{\left(  2\right)  }  &  =\tfrac{e_{-}}{\left(  r+m_{-}\right)
^{2}}\left(  u_{a}\beta^{a}-2\right)  \left[  V\left(  r^{2}-c^{2}\right)
d\Omega_{2}-dr\wedge\left(  d\psi+\omega^{0}\right)  \right] \label{H_1}\\
\tilde{H}_{a}^{\left(  2\right)  }  &  =\left[  \tfrac{e_{-}}{\left(
r+m_{-}\right)  ^{2}}u_{a}\beta^{1}+\tfrac{1}{2}\eta_{ab}\beta^{b}%
Z_{1}\right]  \left[  V\left(  r^{2}-c^{2}\right)  d\Omega_{2}-dr\wedge\left(
d\psi+\omega^{0}\right)  \right] \label{H_a}\\
&  +\tfrac{1}{2}\eta_{ab}\beta^{b}Z_{2}\left[  V\sin\theta dr\wedge
d\phi+d\theta\wedge\left(  d\psi+\omega^{0}\right)  \right] \nonumber\\
\tilde{\Lambda}_{1}  &  =\tfrac{1}{L_{c}L^{c}}\left(  \beta^{a}L_{a}-1\right)
\left(  dt+\hat{k}\right)  -\tfrac{e_{-}}{2m_{-}V}l_{a}\beta^{a}\left(
d\psi+\omega^{0}\right) \label{lambda_big_1}\\
\tilde{\Lambda}_{a}  &  =\tfrac{1}{2}\left(  \tfrac{1}{L^{a}}\beta^{1}%
+\tfrac{1}{L}\eta_{ab}\beta^{b}-\tfrac{1}{LL^{a}}\right)  \left(  dt+\hat
{k}\right)  -\tfrac{e_{-}}{2m_{-}V}l_{a}\beta^{1}\left(  d\psi+\omega
^{0}\right)  . \label{lambda_big_a}%
\end{align}

Note: The tilde indicates that the $d\tilde{\Lambda}_{I}$ are not exact, since
the $\tilde{\Lambda}_{I}$ do not vanish at infinity and are thus singular
there; exactness can be achieved, however, by extracting the cohomological
bits from the $d\tilde{\Lambda}_{I}$ by means of the cohomological basis,
which we will derive later, and shift them into the $\tilde{H}_{I}^{\left(
2\right)  }$.

The $\tilde{H}_{I}^{\left(  2\right)  }$ are manifestly self-dual, and one can
check that $d\tilde{H}_{I}^{\left(  2\right)  }=0$. They are thus harmonic and
can locally be written as $\tilde{H}_{I}^{\left(  2\right)  }=d\tilde{B}_{I}$,
where the $\tilde{B}_{I}$ are not globally defined since they do not vanish at
the pinch-off points of the $\psi$-coordinate:%
\begin{align}
\tilde{B}_{1}  &  =e_{-}\left(  u_{a}\beta^{a}-2\right)  \left[  \tfrac
{1}{r+m_{-}}\left(  d\psi+\omega^{0}\right)  -\left(  \tfrac{\cos\theta}%
{2}+\tfrac{k}{R+m_{-}}\tfrac{r-R\cos\theta}{\sqrt{r_{1}^{2}-c^{2}\sin
^{2}\theta}}\right)  d\phi\right] \label{B_1}\\
\tilde{B}_{a}  &  =e_{-}u_{a}\beta^{1}\left[  \tfrac{1}{r+m_{-}}\left(
d\psi+\omega^{0}\right)  -\left(  \tfrac{\cos\theta}{2}+\tfrac{k}{R+m_{-}%
}\tfrac{r-R\cos\theta}{\sqrt{r_{1}^{2}-c^{2}\sin^{2}\theta}}\right)
d\phi\right] \nonumber\\
&  -\tfrac{1}{4e_{-}}\eta_{ab}\beta^{b}\left[  \tfrac{\left(  r+m_{-}\right)
\left(  c^{2}+m_{-}r\right)  }{V\left(  r^{2}-c^{2}\right)  }\left(
d\psi+\omega^{0}\right)  +\left(  c^{2}-m_{-}^{2}\right)  \cos\theta
d\phi\right]  . \label{B_a}%
\end{align}

Note that, because $V$ has a singularity at each center, $V^{-1}$ goes zero
there and ensures finite norms -- except for the terms with the additional
factor $\left(  r-c\right)  ^{-1}$ which cancels the zero at the bolt and
hence the \textquotedblleft good\textquotedblright\ behavior.

As one can clearly see from the analysis, the factors in the volume integral
of $\left(  \ref{Komar mass}\right)  $, $\tilde{H}_{I}^{\left(  2\right)  }$
and $F^{I}$, are each equipped with two dual flux terms, $V\left(  r^{2}%
-c^{2}\right)  d\Omega_{2}$ and $dr\wedge\left(  d\psi+\omega^{0}\right)  $,
wedge-multiplying to space's volume form. Hence, homology allows for the
existence of purely topological terms in the mass formula, which cannot be
converted into boundary terms, and even has self-intersection, as we will see
in the following.

\subsection{Topological data and intersection homology}

In this subsection, we explicitly derive the topological ingredients flowing
into the formula for the mass and charges. The goal is to see if and how each
of the 2-cycle's contribution goes into either, so that we can precisely make
out their sources of topology in virtue of intersection homology -- with
special emphasis on the breaking of supersymmetry.

As mentioned earlier, the cohomological fluxes $\left(  \ref{H_1}\right)
-\left(  \ref{H_a}\right)  $ are still missing cohomology from the
$d\tilde{\Lambda}_{I}$. Since the $\tilde{\Lambda}_{I}$ are only irregular at
infinity, these cohomological pieces just concern the non-compact cycle,
$C_{\subset}$:%
\begin{align}
H_{1}^{\left(  2\right)  }  &  =\tilde{H}_{1}^{\left(  2\right)  }+\left(
\int_{C_{\subset}}d\tilde{\Lambda}_{1}\right)  v^{\subset}\nonumber\\
&  =\tilde{H}_{1}^{\left(  2\right)  }+\pi k\left[  \left(  2\beta^{a}%
-\tfrac{1}{l_{a}+u_{a}}\right)  \tfrac{\gamma}{l^{a}+u^{a}}-\tfrac{4e_{-}%
}{m_{-}}l_{a}\beta^{a}\right]  v^{\subset}\\
H_{a}^{\left(  2\right)  }  &  =\tilde{H}_{a}^{\left(  2\right)  }+\left(
\int_{C_{\subset}}d\tilde{\Lambda}_{a}\right)  v^{\subset}\nonumber\\
&  =\tilde{H}_{a}^{\left(  2\right)  }+2\pi k\left[  \left(  \tfrac{\gamma
}{l^{a}+u^{a}}-\tfrac{2e_{-}}{m_{-}}l_{a}\right)  \beta^{1}+\tfrac{\gamma
}{\hat{L}}\eta_{ab}\beta^{b}-\tfrac{\gamma}{\hat{L}\left(  l^{a}+u^{a}\right)
}\right]  v^{\subset},
\end{align}
where $v^{\subset}$ is the cohomological basis vector for the non-compact
cycle derived below in $\left(  \ref{v^Non-compact}\right)  $.

The Komar mass and charges are not dependent on any choice of gauge, but it is
very convenient to choose the $\beta^{I}$ such that the boundary integral term
$2\lambda^{I}G_{I}$ in $\left(  \ref{M_rewritten}\right)  $ vanishes; with
$\left(  \ref{lambda_small^1}\right)  $ and $\left(  \ref{lambda_small^a}%
\right)  $ this is achieved by%
\begin{align}
\beta^{1}  &  =\lim_{r\rightarrow\infty}\tfrac{1}{L}=\tfrac{1}{\hat{L}%
}=\left[  \tfrac{e_{-}^{2}}{m_{-}^{2}}l_{a}l^{a}+\tfrac{e_{-}^{2}}%
{c^{2}\left(  c+m_{-}\right)  }\left(  m_{+}\tfrac{R-c}{R+m_{-}}-\tfrac
{4c^{2}k}{\left(  c+m_{-}\right)  ^{2}}\right)  u_{a}u^{a}\right]
^{-1}\label{beta choice_1}\\
\beta^{a}  &  =\lim_{r\rightarrow\infty}\tfrac{1}{L_{a}}=\tfrac{1}{l_{a}%
+u_{a}}. \label{beta choice_a}%
\end{align}

With this choice the equations $\left(  \ref{Komar mass}\right)  $ and
$\left(  \ref{Charge-formula}\right)  $ then become completely topological
expressions:%
\begin{align}
M  &  =-\tfrac{1}{16\pi G_{5}}\int_{\Sigma_{4}}H_{I}^{\left(  2\right)
}\wedge F^{I}\label{M_rewritten}\\
Q^{I}  &  =-\tfrac{1}{64\pi^{2}k}C_{IJK}\int_{\Sigma_{4}}F^{J}\wedge F^{K}.
\label{Q_rewritten}%
\end{align}

We now compute the topological \textquotedblleft building
blocks\textquotedblright\ of the fields and fluxes by which $\left(
\ref{M_rewritten}\right)  -\left(  \ref{Q_rewritten}\right)  $ shall be
represented in the framework of intersection homology. Like in the last
section, we derive the period-integrals by doing the integrals of the 2-form
fluxes over all topologically relevant 2-cycles (see fig. 2):%
\begin{align*}
&
\begin{tabular}
[c]{|l|l|l|}\hline
& $C_{B}=S_{r=c}^{2}$ (Bolt) & $C_{\Delta}=S^{1}\times\left[  c,R\right]  $
(Bubble)\\\hline
$\int F^{1}$ & $\tfrac{2\pi\left(  c^{2}-m_{-}^{2}\right)  +\pi\left(
c+m_{-}\right)  ^{2}n}{2e_{-}}$ & $\tfrac{\pi\left(  c+m_{-}\right)  ^{2}%
}{e_{-}}$\\\hline
$\int F^{a}$ & $\tfrac{16\pi cke_{-}}{\left(  c+m_{-}\right)  ^{2}}u^{a}$ &
$\tfrac{8\pi ke_{-}\left(  c-R\right)  }{\left(  c+m_{-}\right)  \left(
R+m_{-}\right)  }u^{a}$\\\hline
$\int H_{1}^{\left(  2\right)  }$ & $-\tfrac{8\pi cke_{-}}{\left(
c+m_{-}\right)  ^{2}}l_{a}\frac{1}{l_{a}+u_{a}}$ & $-\tfrac{4\pi ke_{-}\left(
c-R\right)  }{\left(  c+m_{-}\right)  \left(  R+m_{-}\right)  }l_{a}\frac
{1}{l_{a}+u_{a}}$\\\hline
$\int H_{a}^{\left(  2\right)  }$ & $\tfrac{8\pi cke_{-}}{\left(
c+m_{-}\right)  ^{2}}\frac{1}{\hat{L}}u_{a}+\tfrac{2\pi\left(  c^{2}-m_{-}%
^{2}\right)  +\pi\left(  c+m_{-}\right)  ^{2}n}{2e_{-}}\frac{1}{l^{a}+u^{a}}$
& $\tfrac{4\pi ke_{-}\left(  c-R\right)  }{\left(  c+m_{-}\right)  \left(
R+m_{-}\right)  }\tfrac{1}{\hat{L}}u_{a}+\tfrac{\pi\left(  c+m\right)  ^{2}%
}{2e_{-}}\frac{1}{l^{a}+u^{a}}$\\\hline
\end{tabular}
\\
&  \text{Table 2}\colon\text{ Integrals of the 2-forms over bolt and bubble
cycle}\\
&
\begin{tabular}
[c]{|l|l|}\hline
& $C_{\subset}=S^{1}\times\left[  R,\infty\right[  $ (Non-compact
cycle)\\\hline
$\int F^{1}$ & $4\pi k\left(  \tfrac{\gamma}{\hat{L}}-\tfrac{2m_{-}}{e_{-}%
}\right)  $\\\hline
$\int F^{a}$ & $4\pi k\left[  \tfrac{\gamma}{l_{a}+u_{a}}-\tfrac{2e_{-}}%
{m_{-}}\left(  l^{a}+\tfrac{m_{-}}{R+m_{-}}u^{a}\right)  \right]  $\\\hline
$\int H_{1}^{\left(  2\right)  }$ & $\pi k\left(  \tfrac{\gamma}{l^{a}+u^{a}%
}-\tfrac{e_{-}}{m_{-}}\tfrac{4R}{R+m_{-}}l_{a}\right)  \tfrac{1}{l_{a}+u_{a}}%
$\\\hline
$\int H_{a}^{\left(  2\right)  }$ & $2\pi k\left[  \left(  \frac{\gamma}%
{\hat{L}}-\tfrac{2m_{-}}{e_{-}}\right)  \tfrac{1}{l^{a}+u^{a}}-\tfrac{2e_{-}%
}{m_{-}}\left(  l_{a}+\tfrac{m_{-}}{R+m_{-}}u_{a}\right)  \frac{1}{\hat{L}%
}\right]  $\\\hline
\end{tabular}
\\
&  \text{Table 3}\colon\text{ Integrals of the 2-forms over the non-compact
cycle}%
\end{align*}
The parameter $n$ hereby represents the bolt's NUT-charge.

It is instructive again to introduce a canonical integer basis for the
cohomology,%
\begin{equation}
\int_{C_{A}}v^{A^{\prime}}=\delta_{A}^{A^{\prime}}\text{ and }\int_{\Sigma
_{4}}v^{A}\wedge v^{A^{\prime}}=I^{AA^{\prime}}, \label{Basis_condition}%
\end{equation}
with%
\begin{equation}
F^{I}=\sigma_{A}^{I}v^{A}\text{ and }H_{I}^{\left(  2\right)  }=\tilde{\sigma
}_{I,A}v^{A}, \label{F_H_composed}%
\end{equation}
where $\sigma_{A}^{I}$ and $\tilde{\sigma}_{I,A}$ ($A=B,\Delta,\subset$)
represent the entries of the above tables, and $I^{AA^{\prime}}=I^{A^{\prime
}A}$ is the inverse intersection matrix.

In this spirit, the bulk integral $\left(  \ref{M_rewritten}\right)  $ becomes%
\begin{equation}
\int_{\Sigma_{4}}H_{I}^{\left(  2\right)  }\wedge F^{I}=\tilde{\sigma}%
_{I,A}\sigma_{A^{\prime}}^{I}I^{AA^{\prime}}, \label{Intersection_integral}%
\end{equation}
and analogously for $\left(  \ref{Q_rewritten}\right)  $. Reproducing the
integrals by composing the products of the period-integrals, $\sigma_{A}^{I}$
and $\tilde{\sigma}_{I,A}$, by the integer coefficients of $I^{AA^{\prime}}$
in the sense of $\left(  \ref{Intersection_integral}\right)  $, can only be
achieved by%
\begin{equation}
I^{AA^{\prime}}=\left(
\begin{array}
[c]{ccc}%
0 & 1 & 1\\
1 & -n & -n\\
1 & -n & -n-1
\end{array}
\right)  \Leftrightarrow I_{AA^{\prime}}=\left(
\begin{array}
[c]{ccc}%
n & 1 & 0\\
1 & -1 & 1\\
0 & 1 & -1
\end{array}
\right)  . \label{Intersection matrix}%
\end{equation}

It is obvious that in the present spacetime the homological structure is
significantly more complex than in the one of the last section.

The off-diagonal 1's in $I_{AA^{\prime}}$ are directly clear: The bolt
intersects the bubble at the former's north pole; the bubble intersects the
non-compact cycle in the Gibbons-Hawking point. Less intuitive are the
self-intersections of each cycle -- it is $n$-fold for the bolt for NUT-charge
of $n$; the self-intersections of the bubble and the non-compact cycle arise,
like the intersection between them, from the topology of the nut linking these cycles.

Since we want to study the explicit topological contribution of each
intersecting part later, it is important to see in how far the elements of
$I_{AA^{\prime}}$, representing the intersections, are reflected in
$I^{AA^{\prime}}$, representing the composition of the building blocks. To
shed light on this, we leave the intersection numbers in $I_{AA^{\prime}}$
more general:%
\begin{equation}
I_{AA^{\prime}}=\left(
\begin{array}
[c]{ccc}%
I_{BB} & I_{B\Delta} & 0\\
I_{B\Delta} & I_{\Delta\Delta} & I_{\Delta\subset}\\
0 & I_{\Delta\subset} & I_{\subset\subset}%
\end{array}
\right)  . \label{I_variation}%
\end{equation}
After inversion, we receive%
\begin{equation}
I^{AA^{\prime}}=\tfrac{1}{\det\left(  I_{AA^{\prime}}\right)  }\left(
\begin{array}
[c]{ccc}%
I_{\Delta\Delta}I_{\subset\subset}-I_{\Delta\subset}^{2} & -I_{B\Delta
}I_{\subset\subset} & I_{B\Delta}I_{\Delta\subset}\\
-I_{B\Delta}I_{\subset\subset} & I_{BB}I_{\subset\subset} & -I_{BB}%
I_{\Delta\subset}\\
I_{B\Delta}I_{\Delta\subset} & -I_{BB}I_{\Delta\subset} & I_{BB}%
I_{\Delta\Delta}-I_{B\Delta}^{2}%
\end{array}
\allowbreak\right)  . \label{I_inverse_variation}%
\end{equation}

Note, that this form is merely a schematic \textquotedblleft tracking
device\textquotedblright\ to qualitatively illustrate how the contributions of
intersection from $I_{AA^{\prime}}$ distribute among the non-zero entries of
$I^{AA^{\prime}}$; hence, the determinant, $\det\left(  I_{AA^{\prime}%
}\right)  =1$, and the upper left entry, $I_{\Delta\Delta}I_{\subset\subset
}-I_{\Delta\subset}^{2}=0$ (not contributing to topology), are not of interest
in this regard. So, the matrix can be written as:%
\begin{align}
I^{AA^{\prime}}  &  =\left(
\begin{array}
[c]{ccc}%
0 & -I_{B\Delta}I_{\subset\subset} & I_{B\Delta}I_{\Delta\subset}\\
-I_{B\Delta}I_{\subset\subset} & I_{BB}I_{\subset\subset} & -I_{BB}%
I_{\Delta\subset}\\
I_{B\Delta}I_{\Delta\subset} & -I_{BB}I_{\Delta\subset} & I_{BB}%
I_{\Delta\Delta}-I_{B\Delta}^{2}%
\end{array}
\allowbreak\right) \\
&  =I_{B\Delta}\left(
\begin{array}
[c]{ccc}%
0 & -I_{\subset\subset} & I_{\Delta\subset}\\
-I_{\subset\subset} & 0 & 0\\
I_{\Delta\subset} & 0 & -I_{B\Delta}%
\end{array}
\right)  \allowbreak+I_{BB}\left(
\begin{array}
[c]{ccc}%
0 & 0 & 0\\
0 & I_{\subset\subset} & -I_{\Delta\subset}\\
0 & -I_{\Delta\subset} & I_{\Delta\Delta}%
\end{array}
\right)  .\allowbreak\label{I_inverse_decomposed}%
\end{align}

The last expression indeed gives a clue about how to disentangle homology.

First, $I^{AA^{\prime}}$ decomposes with respect to the bolt-intersections
($I_{BB}$ and $I_{B\Delta}$). The resulting constituents are predominantly
defined by the homology of the subsystem of the bubble and the non-compact
cycle, that is, the topology of the nut; and if this was turned off, then
there would still be a contribution $-I_{B\Delta}^{2}=-1$ from the bolt-bubble intersection.

This gives rise to the conclusion that there might be a possible topological
hierarchy between the bolt and the nut.

In any case, these insights will prove to be very helpful in spectralizing the
topological contributions later.

From $\left(  \ref{F^1}\right)  -\left(  \ref{Z2}\right)  $ and $\left(
\ref{F_H_composed}\right)  -\left(  \ref{Intersection matrix}\right)  $
follows the cohomological basis:%
\begin{align}
v^{B}  &  =\tfrac{1}{4\pi}d\Omega_{2}\label{v^B}\\
v^{\Delta}  &  =\left(  \Xi_{1}^{\Delta}Z_{1}+\tfrac{\Xi_{2}^{\Delta}}{\left(
r+m\right)  ^{2}}\right)  \left[  V\left(  r^{2}-c^{2}\right)  d\Omega
_{2}-dr\wedge\left(  d\psi+\omega^{0}\right)  \right] \nonumber\\
&  +\Xi_{1}^{\Delta}Z_{2}\left[  d\theta\wedge\left(  d\psi+\omega^{0}\right)
+V\sin\theta dr\wedge d\phi\right]  +\Xi_{3}^{\Delta}d\Omega_{2}%
\label{v^Delta}\\
v^{\subset}  &  =\left(  \Xi_{1}^{\subset}Z_{1}+\tfrac{\Xi_{2}^{\subset}%
}{\left(  r+m\right)  ^{2}}\right)  \left[  V\left(  r^{2}-c^{2}\right)
d\Omega_{2}-dr\wedge\left(  d\psi+\omega^{0}\right)  \right] \nonumber\\
&  +\Xi_{1}^{\subset}Z_{2}\left[  d\theta\wedge\left(  d\psi+\omega
^{0}\right)  +V\sin\theta dr\wedge d\phi\right]  +\Xi_{3}^{\subset}d\Omega
_{2}, \label{v^Non-compact}%
\end{align}
with the constant coefficients,%
\begin{equation}%
\begin{array}
[c]{ll}%
\Xi_{1}^{\Delta}=\tfrac{e_{-}}{\pi}\tfrac{c+m_{-}}{\left(  c+m_{-}\right)
^{3}-4km_{-}\left(  c-R\right)  } & \Xi_{1}^{\subset}=\tfrac{e_{-}}{\pi}%
\tfrac{c-R}{\left(  c+m_{-}\right)  ^{3}-4km_{-}\left(  c-R\right)  }\\
\Xi_{2}^{\Delta}=-\tfrac{m_{-}}{\pi}\tfrac{\left(  R+m_{-}\right)  \left(
c+m_{-}\right)  }{\left(  c+m_{-}\right)  ^{3}-4km_{-}\left(  c-R\right)  } &
\Xi_{2}^{\subset}=-\tfrac{1}{4\pi k}\tfrac{\left(  R+m_{-}\right)  \left(
c+m_{-}\right)  ^{3}}{\left(  c+m_{-}\right)  ^{3}-4km_{-}\left(  c-R\right)
}\\
\Xi_{3}^{\Delta}=\tfrac{8ckm_{-}\left(  R+m_{-}\right)  -\left(
c+m_{-}\right)  ^{3}\left[  2\left(  c-m_{-}\right)  +\left(  c+m_{-}\right)
n\right]  }{4\pi\left(  c+m_{-}\right)  \left[  \left(  c+m_{-}\right)
^{3}-4km_{-}\left(  c-R\right)  \right]  } & \Xi_{3}^{\subset}=\tfrac{c+m_{-}%
}{4\pi}\tfrac{\left(  n-2\right)  \left(  R-c\right)  \left(  c+m_{-}\right)
+2c\left(  3R-2c+m_{-}\right)  }{\left(  c+m_{-}\right)  ^{3}-4km_{-}\left(
c-R\right)  }%
\end{array}
\end{equation}

Note: $v^{\Delta}$ and $v^{\subset}$ are each self-dual and hence harmonic, up
to a $d\Omega_{2}$-term.

The orthonormality condition $\left(  \ref{Basis_condition}\right)  $ is
obviously fulfilled by $v^{B}$. However, to show the same for $v^{\Delta}$ and
$v^{\subset}$, one has to evaluate the integrals of the coefficent functions
over the cycles with help of%
\begin{align}
\int_{S_{r=c}^{2}}V\left(  r^{2}-c^{2}\right)  Z_{1}d\Omega_{2}  &
=\tfrac{2\pi\left(  c^{2}-m_{-}^{2}\right)  }{e_{-}}+\tfrac{\pi\left(
c+m_{-}\right)  ^{2}}{e_{-}}n\\
\int_{c}^{R}Z_{1}|_{\theta=0}dr  &  =-\tfrac{\left(  c+m_{-}\right)  ^{2}%
}{4ke_{-}}\\
\int_{R}^{\infty}Z_{1}|_{\theta=0}dr  &  =\tfrac{m_{-}}{e_{-}}.
\end{align}

Now, we can easily write the fluxes in terms of the cohomology basis:%
\begin{align}
H_{1}^{\left(  2\right)  }  &  =\tilde{\sigma}_{1,A}v^{A}\nonumber\\
&  =-\left[  \tfrac{8ce_{-}}{\left(  c+m_{-}\right)  ^{2}}l_{a}v^{B}%
+\tfrac{4e_{-}\left(  c-R\right)  }{\left(  c+m_{-}\right)  \left(
R+m_{-}\right)  }l_{a}v^{\Delta}-\left(  \tfrac{\gamma}{l^{a}+u^{a}}%
-\tfrac{e_{-}}{m_{-}}\tfrac{4R}{R+m_{-}}l_{a}\right)  v^{\subset}\right]
\tfrac{\pi k}{l_{a}+u_{a}}\\
H_{a}^{\left(  2\right)  }  &  =\tilde{\sigma}_{a,A}v^{A}\nonumber\\
&  =\left[  \tfrac{8\pi cke_{-}}{\left(  c+m_{-}\right)  ^{2}}\tfrac{1}%
{\hat{L}}u_{a}+\left(  \tfrac{\pi\left(  c^{2}-m_{-}^{2}\right)  }{e_{-}%
}+\tfrac{\pi\left(  c+m_{-}\right)  ^{2}n}{2e_{-}}\right)  \tfrac{1}%
{l^{a}+u^{a}}\right]  v^{B}\nonumber\\
&  +\left[  \tfrac{4\pi ke_{-}\left(  c-R\right)  }{\left(  c+m_{-}\right)
\left(  R+m_{-}\right)  }\tfrac{1}{\hat{L}}u_{a}+\tfrac{\pi\left(  c+m\right)
^{2}}{2e_{-}}\tfrac{1}{l^{a}+u^{a}}\right]  v^{\Delta}\\
&  -2\pi k\left[  \tfrac{2e_{-}}{m_{-}}\left(  l_{a}+\tfrac{m_{-}}{R+m_{-}%
}u_{a}\right)  \tfrac{1}{\hat{L}}-\left(  \tfrac{\gamma}{\hat{L}}%
-\tfrac{2m_{-}}{e_{-}}\right)  \tfrac{1}{l^{a}+u^{a}}\right]  v^{\subset
}\nonumber\\
\Lambda_{1}  &  =\tilde{\Lambda}_{1}-\left(  \int_{C_{\subset}}d\tilde
{\Lambda}_{1}\right)  \omega^{\subset}\nonumber\\
&  =\tfrac{1}{L_{c}L^{c}}\left(  \tfrac{1}{l_{a}+u_{a}}L_{a}-1\right)  \left(
dt+\hat{k}\right)  -\tfrac{e_{-}}{2m_{-}V}l_{a}\tfrac{1}{l_{a}+u_{a}}\left(
d\psi+\omega^{0}\right)  -\tfrac{\pi k}{l_{a}+u_{a}}\tilde{\gamma}_{a}%
\omega^{\subset}\\
\Lambda_{a}  &  =\tilde{\Lambda}_{a}-\left(  \int_{C_{\subset}}d\tilde
{\Lambda}_{a}\right)  \omega^{\subset}\nonumber\\
&  =\tfrac{1}{2}\left(  \tfrac{1}{L^{a}}\beta^{1}+\tfrac{1}{L}\eta_{ab}%
\beta^{b}-\tfrac{1}{LL^{a}}\right)  \left(  dt+\hat{k}\right)  -\tfrac{e_{-}%
}{2m_{-}V}l_{a}\beta^{1}\left(  d\psi+\omega^{0}\right)  -\tfrac{2\pi k}%
{\hat{L}}\gamma_{a}\omega^{\subset},
\end{align}
where%
\begin{align}
\gamma_{a}  &  =\tfrac{\gamma}{l^{a}+u^{a}}-\tfrac{2e_{-}}{m_{-}}l_{a}\\
\tilde{\gamma}_{a}  &  =\tfrac{\gamma}{l^{a}+u^{a}}-\tfrac{4e_{-}}{m_{-}}%
l_{a},
\end{align}
and $\omega^{\subset}$ is the potential for $v^{\subset}=d\omega^{\subset}$,%
\begin{align}
\omega^{\subset}  &  =\left[  \tfrac{1}{r+m_{-}}\Xi_{2}^{\subset}%
-\tfrac{\left(  r+m_{-}\right)  \left(  c^{2}+m_{-}r\right)  }{2e_{-}V\left(
r^{2}-c^{2}\right)  }\Xi_{1}^{\subset}\right]  \left(  d\psi+\omega^{0}\right)
\nonumber\\
&  -\left[  \tfrac{\cos\theta}{2}\left(  \tfrac{c^{2}-m_{-}^{2}}{e_{-}}\Xi
_{1}^{\subset}+\Xi_{2}^{\subset}+2\Xi_{3}^{\subset}\right)  +\tfrac{k}%
{R+m_{-}}\tfrac{r-R\cos\theta}{\sqrt{r_{1}^{2}-c^{2}\sin^{2}\theta}}\Xi
_{2}^{\subset}\right]  d\phi.
\end{align}

Analogously it holds for the fields,%
\begin{align}
F^{1}  &  =\sigma_{A}^{1}v^{A}\\
&  =\left(  \tfrac{2\pi\left(  c^{2}-m_{-}^{2}\right)  }{e_{-}}+\tfrac
{\pi\left(  c+m_{-}\right)  ^{2}n}{e_{-}}\right)  v^{B}+\tfrac{\pi\left(
c+m_{-}\right)  ^{2}}{e_{-}}v^{\Delta}+4\pi k\left(  \tfrac{\gamma}{\hat{L}%
}-\tfrac{2m_{-}}{e_{-}}\right)  v^{\subset}\\
F^{a}  &  =\sigma_{A}^{a}v^{A}\\
&  =\tfrac{8\pi ke_{-}}{c+m_{-}}u^{a}\left(  \tfrac{2c}{c+m_{-}}v^{B}%
+\tfrac{c-R}{R+m_{-}}v^{\Delta}\right)  +4\pi k\left[  \tfrac{\gamma}%
{l_{a}+u_{a}}-\tfrac{2e_{-}}{m_{-}}\left(  l^{a}+\tfrac{m_{-}}{R+m_{-}}%
u^{a}\right)  \right]  v^{\subset},
\end{align}
and the potentials of the $H_{I}^{\left(  2\right)  }$,%
\begin{align}
B_{1}  &  =\tilde{B}_{1}+\left(  \int_{C_{\subset}}d\tilde{\Lambda}%
_{1}\right)  \omega^{\subset}=\tilde{B}_{1}+\tfrac{\pi k}{l_{a}+u_{a}}%
\tilde{\gamma}_{a}\omega^{\subset}\\
B_{a}  &  =\tilde{B}_{a}+\left(  \int_{C_{\subset}}d\tilde{\Lambda}%
_{a}\right)  \omega^{\subset}=\tilde{B}_{a}+\tfrac{2\pi k}{\hat{L}}\gamma
_{a}\omega^{\subset}.
\end{align}

\subsection{Mass, charges, and BPS-bound breaking from cohomology}

We now have the data we need to compute the total mass and charges.

From $\left(  \ref{M_rewritten}\right)  -\left(  \ref{Intersection matrix}%
\right)  $ follow the expressions:%
\begin{align}
M  &  =-\tfrac{1}{16\pi G_{5}}\int_{\Sigma_{4}}H_{I}^{\left(  2\right)
}\wedge F^{I}=-\tfrac{1}{16\pi G_{5}}\tilde{\sigma}_{I,A}\sigma_{A^{\prime}%
}^{I}I^{AA^{\prime}}\\
&  =\tfrac{\pi k}{2G_{5}}\left\{  \tfrac{4ke_{-}^{2}}{\hat{L}}\left[
\tfrac{1}{\left(  R+m_{-}\right)  ^{2}}+\tfrac{4c+n\left(  c+m_{-}\right)
}{\left(  c+m_{-}\right)  ^{3}}\right]  u_{a}u^{a}\right. \nonumber\\
&  \left.  -\tfrac{1}{l_{a}+u_{a}}\left[  \left(  \tfrac{3}{4}\gamma
_{a}-\tfrac{e}{2m}l_{a}\right)  \left(  \tfrac{e_{-}}{\hat{L}}\tfrac{R+2m_{-}%
}{R+m_{-}}u_{b}u^{b}-\tfrac{c^{2}}{e_{-}}\right)  +\tfrac{e_{-}\gamma}%
{2\hat{L}}\left(  u_{a}-l_{a}\right)  \right]  \right\} \label{Mass_total}\\
Q^{1}  &  =-\tfrac{1}{64\pi^{2}k}\eta_{ab}\int_{\Sigma_{4}}F^{a}\wedge
F^{b}=-\tfrac{1}{64\pi^{2}k}\eta_{ab}\sigma_{A}^{a}\sigma_{A^{\prime}}%
^{b}I^{AA^{\prime}}\\
&  =ke_{-}^{2}\left[  \tfrac{1}{\left(  R+m_{-}\right)  ^{2}}+\tfrac
{4c+n\left(  c+m_{-}\right)  }{\left(  c+m_{-}\right)  ^{3}}\right]
u_{a}u^{a}+\tfrac{1}{4}\gamma^{a}\left(  m_{-}\gamma_{a}-2e_{-}u_{a}\right)
\label{Charge^1_total}\\
Q^{a}  &  =-\tfrac{1}{32\pi^{2}k}\eta_{ab}\int_{\Sigma_{4}}F^{1}\wedge
F^{b}=-\tfrac{1}{32\pi^{2}k}\eta_{ab}\sigma_{A}^{1}\sigma_{A^{\prime}}%
^{b}I^{AA^{\prime}}\\
&  =-\tfrac{1}{4}\left[  \tfrac{2m_{-}\gamma}{\hat{L}}\left(  \tfrac{e_{-}%
}{m_{-}}u_{a}-\gamma_{a}\right)  +\tfrac{m_{-}^{2}-c^{2}}{e_{-}}\gamma
_{a}\right]  . \label{Charge^a_total}%
\end{align}

The core piece of this work is to examine the topological origin of the
BPS-bound breaking extra-mass, for which purpose we will sequence the
foregoing results\ with respect to intersecting homology.

Comparing $\left(  \ref{Mass_total}\right)  $ with $\left(
\ref{Charge^1_total}\right)  -\left(  \ref{Charge^a_total}\right)  $, the
relation between the mass and the total charge,%
\begin{equation}
M=\tfrac{2\pi k}{G_{5}}\left(  Q^{1}+Q^{2}+Q^{3}\right)  +\Delta M,
\label{Mass-charge-relation}%
\end{equation}
gives the sought extra-mass term,%
\begin{align}
\Delta M  &  =M-\tfrac{2\pi k}{G_{5}}\Sigma_{I=1}^{3}Q^{I}\\
&  =-\tfrac{1}{16\pi G_{5}}\left[  \int_{\Sigma_{4}}H_{I}^{\left(  2\right)
}\wedge F^{I}-\tfrac{1}{2}\Sigma_{I=1}^{3}C_{IJK}\int_{\Sigma_{4}}F^{J}\wedge
F^{K}\right] \\
&  =-\tfrac{1}{16\pi G_{5}}\sigma_{A^{\prime}}^{K}\left(  \tilde{\sigma}%
_{K,A}-\tfrac{1}{2}\Sigma_{I=1}^{3}C_{IJK}\sigma_{A}^{J}\right)
I^{AA^{\prime}}. \label{Extra-mass_general}%
\end{align}
With $\left(  \ref{Mass_total}\right)  -\left(  \ref{Charge^a_total}\right)  $
it ammounts to%
\begin{align}
\Delta M  &  =\tfrac{\pi k}{2G_{5}}\Sigma_{a=2}^{3}\left\{  4ke_{-}^{2}\left(
\tfrac{1}{\hat{L}}-1\right)  \left[  \tfrac{1}{\left(  R+m_{-}\right)  ^{2}%
}+\tfrac{4c+n\left(  c+m_{-}\right)  }{\left(  c+m_{-}\right)  ^{3}}\right]
u_{a}u^{a}\right. \nonumber\\
&  -\tfrac{1}{l_{a}+u_{a}}\left(  \tfrac{3}{4}\gamma_{a}-\tfrac{e}{2m}%
l_{a}\right)  \left(  \tfrac{e_{-}}{\hat{L}}\tfrac{R+2m_{-}}{R+m_{-}}%
u_{b}u^{b}-\tfrac{c^{2}}{e_{-}}\right)  -\tfrac{e_{-}\gamma}{\hat{L}}\left[
u_{a}\left(  \tfrac{1}{l_{a}+u_{a}}-1\right)  -1\right]
\label{Extra-mass_computed}\\
&  \left.  +\left(  \tfrac{\gamma}{2\hat{L}}+\gamma^{a}\right)  \left(
2e_{-}u_{a}-m_{-}\gamma_{a}\right)  +\left(  \tfrac{m_{-}^{2}-c^{2}}{e_{-}%
}-\tfrac{3m_{-}\gamma}{2\hat{L}}\right)  \gamma_{a}\right\}  .\nonumber
\end{align}

Like in the previous section, it is insightful to investigate the from
$\left(  \ref{Extra-mass_general}\right)  $ factored term,%
\begin{equation}
\chi_{K,A}=\tilde{\sigma}_{K,A}-\tfrac{1}{2}\Sigma_{I=1}^{3}C_{IJK}\sigma
_{A}^{J}, \label{Extra-mas_factored_term}%
\end{equation}
in more detail for every cycle:%
\begin{align}
\chi_{1,B}  &  =-\tfrac{8\pi cke_{-}}{\left(  c+m_{-}\right)  ^{2}}\left[
2-\Sigma_{a=2}^{3}u_{a}\left(  \tfrac{1}{l_{a}+u_{a}}-1\right)  \right]
\label{factor1}\\
\chi_{a,B}  &  =\tfrac{8\pi cke_{-}}{\left(  c+m_{-}\right)  ^{2}}\left(
\tfrac{1}{\hat{L}}-1\right)  u_{a}+\left(  \tfrac{\pi\left(  c^{2}-m_{-}%
^{2}\right)  }{e_{-}}+\tfrac{\pi\left(  c+m_{-}\right)  ^{2}}{2e_{-}}n\right)
\left(  \tfrac{1}{l^{a}+u^{a}}-1\right) \\
\chi_{1,\Delta}  &  =-\tfrac{4\pi ke_{-}\left(  c-R\right)  }{\left(
c+m_{-}\right)  \left(  R+m_{-}\right)  }\left[  2-\Sigma_{a=2}^{3}%
u_{a}\left(  \tfrac{1}{l_{a}+u_{a}}-1\right)  \right] \\
\chi_{a,\Delta}  &  =\tfrac{4\pi ke_{-}\left(  c-R\right)  }{\left(
c+m_{-}\right)  \left(  R+m_{-}\right)  }\left(  \tfrac{1}{\hat{L}}-1\right)
u_{a}+\tfrac{\pi\left(  c+m\right)  ^{2}}{2e_{-}}\left(  \tfrac{1}{l^{a}%
+u^{a}}-1\right) \\
\chi_{1,\subset}  &  =\pi k\Sigma_{a=2}^{3}\left\{  \left[  \tfrac{\gamma
}{l^{a}+u^{a}}-\tfrac{4e_{-}}{m_{-}}\left(  l_{a}+\tfrac{m_{-}}{R+m_{-}}%
u_{a}\right)  \right]  \left(  \tfrac{1}{l_{a}+u_{a}}-1\right)  -\tfrac
{\gamma}{l^{a}+u^{a}}+\tfrac{4e_{-}}{R+m_{-}}\right\} \\
\chi_{a,\subset}  &  =2\pi k\left[  \left(  \tfrac{\gamma}{\hat{L}}%
-\tfrac{2m_{-}}{e_{-}}\right)  \left(  \tfrac{1}{l^{a}+u^{a}}-1\right)
-\tfrac{2e_{-}}{m_{-}}\left(  l_{a}+\tfrac{m_{-}}{R+m_{-}}u_{a}\right)
\left(  \tfrac{1}{\hat{L}}-1\right)  -\tfrac{\gamma}{l^{a}+u^{a}}\right]  .
\label{factor6}%
\end{align}

In the previous section, all contribution from the bolt canceled out
identically, so that only the non-compact cycle turned out to be responsible
for the breaking of supersymmetry. Here we can clearly see that every cycle
contributes, if one keeps the parameters general.

In the following, we consider a special choice of parameters, in order to both
simplify the foregoing results and parallel the procedure even more with the
one from the previous section.

In section two, the warp factor of the five-dimensional metric, $Z$, goes to 1
at infinity. For the warp factor, $\frac{1}{2}LL_{a}L^{a}$, of the present
spacetime, however, this is not so obvious, since the asymptotics of $L_{a}$
and $L$ are composed of the very strictly bound parameters (see appendix B).
Following the regularity discussion at the end of \cite{BubblingBolt}, one
condition outlined (eq. 4.17) is%
\begin{equation}
VL>0\text{ and }VL_{a}>0\text{ everywhere,}%
\end{equation}
and so, with $V\rightarrow\frac{1}{2}$ at infinity, we learn that both $L$ and
$L_{a}$ must have positive asymptotics. Hence, the choice%
\begin{equation}
\hat{L}=1\text{ and }l_{a}+u_{a}=l^{a}+u^{a}=l+u=1\text{,} \label{L_choice}%
\end{equation}
is in agreement with that, and doing so we would have%
\begin{equation}
\tfrac{1}{2}LL_{a}L^{a}\rightarrow\tfrac{1}{2}\hat{L}\left(  l_{a}%
+u_{a}\right)  \left(  l^{a}+u^{a}\right)  =\hat{L}\left(  l+u\right)  ^{2}=1,
\end{equation}
as desired.

This way $\left(  \ref{Extra-mass_computed}\right)  $ simplifies significantly
to%
\begin{equation}
\Delta M=\tfrac{\pi ke_{-}\gamma}{2G_{5}}\left(  2l+5-\tfrac{3m_{-}\gamma
}{e_{-}}-\tfrac{c^{2}-m_{-}^{2}}{2e_{-}^{2}}\right)  ,
\end{equation}
where redefining the (restricted) degree of freedom $l>0$ like $2l+5=\frac
{2m_{-}}{e_{-}}\alpha$ yields%
\begin{equation}
\Delta M=\tfrac{\pi km_{-}\gamma}{G_{5}}\left(  \alpha-\tfrac{3\gamma}%
{2}-\tfrac{c^{2}-m_{-}^{2}}{4m_{-}e_{-}}\right)  .
\end{equation}
This result compares nicely to $\left(  \ref{Extra-mass_running-bolt}\right)
$ from the running-bolt spacetime, especially if one does%
\begin{equation}
m_{-}\rightarrow m_{\text{Schwarzschild}}\text{ and }k\left(  \alpha
-\tfrac{3\gamma}{2}-\tfrac{c^{2}-m_{-}^{2}}{4m_{-}e_{-}}\right)
\rightarrow\varepsilon\Sigma_{I=1}^{3}q_{I}.
\end{equation}

The choice $\left(  \ref{L_choice}\right)  $ applied to $\left(
\ref{factor1}\right)  -\left(  \ref{factor6}\right)  $, yields furthermore%
\begin{equation}%
\begin{array}
[c]{lll}%
\chi_{1,B}=-\tfrac{16\pi cke_{-}}{\left(  c+m_{-}\right)  ^{2}} &
\chi_{1,\Delta}=-\tfrac{8\pi ke_{-}\left(  c-R\right)  }{\left(
c+m_{-}\right)  \left(  R+m_{-}\right)  } & \chi_{1,\subset}=-2\pi k\left(
\gamma-\tfrac{4e_{-}}{R+m_{-}}\right) \\
\chi_{a,B}=0 & \chi_{a,\Delta}=0 & \chi_{a,\subset}=-2\pi k\gamma
\end{array}
,
\end{equation}
so clearly every cycle goes at least in part into the extra-mass and so
contributes to the violation of the BPS-bound, where, like in the previous
section, the non-compact cycle yields the strongest contribution.

To see the cycles' contribution more particularly in this light, we decompose
$\Delta M\propto\sigma_{A}^{K}\chi_{K,A^{\prime}}I^{AA^{\prime}}$ into a part
where the compact cycles only \textquotedblleft talk\textquotedblright\ to one
another (c-c) ($A,A^{\prime}=B,\Delta$), and one where the compact cycles
correspond with the non-compact cycle (c-n):%
\begin{align}
\Delta M_{\text{c-c}}  &  =-\tfrac{1}{16\pi G_{5}}\left[  \sigma_{\Delta}%
^{1}\chi_{1,B}+\left(  \sigma_{B}^{1}-n\sigma_{\Delta}^{1}\right)
\chi_{1,\Delta}\right] \nonumber\\
&  =\tfrac{\pi k}{G_{5}}\tfrac{c^{2}+Rm_{-}}{R+m_{-}}\\
\Delta M_{\text{c-n}}  &  =-\tfrac{1}{16\pi G_{5}}\left[  \left(  \sigma
_{B}^{K}-n\sigma_{\Delta}^{K}-\left(  n+1\right)  \sigma_{\subset}^{K}\right)
\chi_{K,\subset}+\sigma_{\subset}^{1}\left(  \chi_{1,B}-n\chi_{1,\Delta
}\right)  \right] \nonumber\\
&  =\tfrac{\pi k}{G_{5}}\left[  \gamma\left(  \tfrac{\left(  2\alpha
-3\gamma\right)  m_{-}}{2}-\tfrac{c^{2}-m_{-}^{2}}{4e_{-}}\right)
-\tfrac{c^{2}+Rm_{-}}{R+m_{-}}\right]  .
\end{align}
So, there is a non-vanishing contribution from the compact cycles only.

Now, we take a closer look at how these results can be interpreted in the
language of intersection homology. The period-integrals compose the extra-mass
integral $\left(  \ref{Extra-mass_general}\right)  $ through the inverse
matrix, $I^{AA^{\prime}}$; from $\left(  \ref{I_inverse_decomposed}\right)  $
we see which cycles intersect in each component.

One finds from $\Delta M\propto\chi_{K,A}\sigma_{A^{\prime}}^{K}I^{AA^{\prime
}}$ and $\left(  \ref{I_inverse_decomposed}\right)  $ that the vanishing of
the components, $\chi_{a,B}$ and $\chi_{a,\Delta}$, weakens the contributions
from the intersection terms: $I_{B\Delta}I_{\subset\subset}$, $I_{B\Delta
}I_{\Delta\subset}$, $I_{BB}I_{\subset\subset}$, and $I_{BB}I_{\Delta\subset}%
$; so, all intersections are affected equally many times -- except for the
self-intersection of the bubble, $I_{\Delta\Delta}$. On the other hand, the
above-mentioned dominance of the non-compact cycle means with $\left(
\ref{I_inverse_decomposed}\right)  $ a stronger representation of the
intersection terms: $I_{B\Delta}I_{\Delta\subset}$, $I_{B\Delta}^{2}$,
$I_{BB}I_{\Delta\subset}$, and $I_{BB}I_{\Delta\Delta}$; this time being all
intersections except for the self-intersection of the non-compact cycle,
$I_{\subset\subset}$. So, under the bottom line one can make a qualitative
estimation of the order of contribution strength for the intersections
(starting with the strongest):%
\begin{equation}
I_{B\Delta}\rightarrow\left(  I_{BB},I_{\Delta\subset}\right)  \rightarrow
\left(  I_{\Delta\Delta},I_{\subset\subset}\right)  .
\end{equation}
Since in the previous section the only contributing (and existing)
intersection was $I_{B\subset}$, one may rise the question whether the
intersection of the bolt with its adjacent cycle might be generally the
dominant one.

In any case, it is striking that the breaking of supersymmetry has for this
solution of supergravity non-vanishing topological contributions from all
cycle's intersections.

\section{Conclusion}

In the long line of efforts in deriving and applying the Smarr formula in a
huge and manifold framework of physical situations, the so-called
\textquotedblleft no-go theorems\textquotedblright, which prohibit the
existence of massive supergravity solitons in the abscence of singularities
and horizons, were shown to be circumvented in recent works by allowing
non-trivial topology to spacetimes of dimensions five or higher.

In this work, further accomplishments on exploring the scope of applicability
and implications of Smarr's formula have been demonstrated and discussed
within the cases of two geometrically and topologically distinct spacetimes.

Since the known non-BPS solutions are very specialized, the assumption stood
to reason that the breaking of supersymmetry might be in general a boundary
phenomenon rather than an intrinsic part of the core solution. In this work,
it could be shown that the breaking of supersymmetry gets indeed contributions
from the core topology in addition to the boundary.

First, we have derived the Komar mass for an non-BPS solution of supergravity
in a five-dimensional stationary spacetime where we gave space a magnetically
charged \textquotedblleft bolt\textquotedblright\ at the center and made it
asymptotically $S^{1}\times%
\mathbb{R}
^{3}$.

The very goal was to determine explicitely how each mass component follows
from topology and especially how cohomology accounts particularly for the
extra-mass causing the violation of the BPS-bound.

One essential question addressed was, whether the extra-mass term violating
the BPS-bound in spacetimes asymptotically behaving like $%
\mathbb{R}
^{1,3}\times S^{1}$, only appears in the ADM mass while the Komar mass
preserves the BPS formula.

It was shown that for a vast simplification of the running bolt solution, in
which the magnetic charges got turned off, this still holds; but in the more
general situation, the Komar mass contains terms breaking supersymmetry as well.

The mass formula is all topology. This is due to the fact that the present
spacetime allows for two harmonic fluxes -- the volume form of the bolt and
its dual flux on the non-compact cycle. Nontheless, the topology of the
Euclidian Schwarzschild base space does not inhabit any self-intersecting
homology as opposed to the Gibbons-Hawking base and the spacetime in second two.

In any case, it is the fluxes on the non-compact cycle that render the
supersymmetry-breaking extra-mass term non-zero, which raised the question
whether the latter is in general a boundary term.

A whole different picture arose when the cohomological fluxes of a 2-center
solution of five-dimensional supergravity were derived, which consists of a
non-extremal magnetic bolt and an extremal Gibbons-Hawking center, both linked
by a bubble 2-cycle.

A first striking consequence from the in this light computed 2-form harmonics
is that the 2-center situation exhibits additional topological flux through
the bubble's pinching off at two centers and thus generating a kind of
\textquotedblleft interaction flux\textquotedblright, which is dual to the
bolt-flux. As opposed to the 1-center running-bolt\ solution, the homology of
this solution turned out to be self-intersecting and the purely topological
contributions to the asymptotic mass and Maxwell-charges much more various.
From the running-bolt solution it is already known that Komar does indeed
reflect the breaking of supersymmetry in a spacetime with the given
asymptotics; the main question, however, addressed at this point is, in which
explicit manner it is caused by the various given features of space's topology.

From the derived harmonic\ fluxes, the formulae for the mass and charges were
computed and the topological pieces of the BPS-bound breaking extra-mass term
explicitly analyzed in virtue of intersection homology. As a striking result,
the extra-mass and hence the breaking of supersymmetry are indeed supported by
all existing intersections of the homological cycles -- dominated by the
bolt-bubble intersection --, so in part even by the compact cycles and hence
the core solution alone.

\bigskip\medskip

\leftline{\bf Acknowledgements}My gratitude goes to Nicholas Warner for his
great work as my doctoral adviser. Further thanks shall be directed toward the
High Energy Group of USC and the relaxing, work-supportive atmosphere of the
University of Southern California. Last but not least would I like to thank
DOE grant DE-SC0011687 by whom this work was in part supported.

\appendix

\section{Functions and constants}

In section 3, a more general solution of supergravity has been considered. It
contains several degrees of freedom, that are non-trivially interrelated, and
some fairly bulky functions. Although the latter were already written out in
section 3, they shall be briefly listed here again for completeness.

The functions used here are taken from eqs. $\left(  2.25\right)  $, $\left(
2.26\right)  $, $\left(  3.55\right)  $, $\left(  4.2\right)  -\left(
4.6\right)  $ of \cite{BubblingBolt}:%
\begin{align}
V\left(  r,\theta\right)   &  =\tfrac{r+m_{-}}{2\left(  r^{2}-c^{2}\right)
}\left(  r+m_{+}-\tfrac{2k}{R+m_{-}}\tfrac{Rr-c^{2}\cos\theta}{\sqrt
{r^{2}+R^{2}-2Rr\cos\theta-c^{2}\sin^{2}\theta}}\right) \label{V_full}\\
\omega^{0}\left(  r,\theta\right)   &  =-\tfrac{1}{2}\left[  \left(
m_{+}-m_{-}\right)  \cos\theta+\tfrac{2k}{R+m_{-}}\tfrac{R^{2}-m_{-}r-R\left(
r-m_{-}\right)  \cos\theta-c^{2}\sin^{2}\theta}{\sqrt{r^{2}+R^{2}%
-2Rr\cos\theta-c^{2}\sin^{2}\theta}}\right]  d\phi\\
\omega\left(  r,\theta\right)   &  =-\tfrac{e_{-}R}{2\left(  R+m_{-}\right)
^{2}}u_{a}u^{a}\left[  \left(  1-\tfrac{r+R}{\sqrt{r^{2}-2Rr\cos\theta+R^{2}}%
}\right)  \left(  1-\cos\theta\right)  +\tfrac{c^{2}}{R\sqrt{r^{2}%
-2Rr\cos\theta+R^{2}}}\sin^{2}\theta\right]  d\phi\\
\hat{k}\left(  r,\theta\right)   &  =\omega-\tfrac{M}{V}\left(  d\psi
+\omega^{0}\right) \\
L_{a}\left(  r,\theta\right)   &  =\tfrac{\left(  r+m_{-}\right)  \left(
c^{2}+m_{-}r\right)  }{2m_{-}\left(  r^{2}-c^{2}\right)  }\tfrac{l_{a}}%
{V}+u_{a}\\
L\left(  r,\theta\right)   &  =\tfrac{e_{-}^{2}}{2m_{-}^{2}}\tfrac{1}{V}%
l_{a}l^{a}-\tfrac{e_{-}^{2}}{c^{2}\left(  c+m_{-}\right)  ^{2}}\tfrac
{f_{1}r+f_{2}}{\left(  m_{-}+r\right)  \left(  m_{-}+R\right)  }u_{a}u^{a}\\
M\left(  r,\theta\right)   &  =-\tfrac{e_{-}}{2m_{-}}l_{a}L^{a}+\tfrac{e_{-}%
}{2\left(  m_{-}+R\right)  }\left[  \tfrac{R-r}{m_{-}+r}V+\tfrac{\left(
c^{2}+m_{-}r\right)  \left(  f_{1}r+f_{2}\right)  }{2c^{2}\left(
c+m_{-}\right)  ^{2}\left(  r^{2}-c^{2}\right)  }\right]  u_{a}u^{a}.
\end{align}
$\allowbreak$

The constants $e_{\pm}$, $m_{\pm}$, $l^{a}$, $u^{a}$, $k$, $f_{1}$, $f_{2}$,
and the NUT-charge, $n$, are connected by the relations:
\begin{align}
c^{2}  &  =m_{+}m_{-}-2e_{+}e_{-}\\
m_{+}  &  =c\left(  -1+\tfrac{4k}{c+m_{-}}+\tfrac{2k}{R+m_{-}}\right)
=\tfrac{2kc}{c+m_{-}}\left(  \tfrac{2m_{-}}{c+m_{-}}-n\right) \\
R  &  =-m_{-}+\tfrac{2k\left(  c+m_{-}\right)  ^{2}}{\left(  c+m_{-}\right)
^{2}-4ck-2k\left(  c+m_{-}\right)  n}\\
-\left(  n+1\right)  k  &  =\tfrac{m_{+}-m_{-}}{2}-\tfrac{kR}{R+m_{-}}\\
f_{1}  &  =\tfrac{m_{+}\left(  c+m\right)  ^{2}\left(  c-R\right)
+4c^{2}k\left(  R+m\right)  }{c+m}\\
f_{2}  &  =\tfrac{cm_{+}\left(  c+m\right)  ^{2}\left(  R-c\right)
+4c^{2}k\left(  c^{2}-2Rc-Rm\right)  }{c+m}. \label{f2_full}%
\end{align}

\section{Asymptotic limits}

\subsection{Functions}

The asymptotic limits of the functions at the bolt and the Gibbons-Hawking
center are:%
\begin{align*}
&
\begin{tabular}
[c]{|c|c|c|}\hline
& $r\rightarrow c$ & $\theta=0,r\rightarrow R$\\\hline
$V$ & $\tfrac{k}{r-c}$ & $-\tfrac{k}{\left\vert R-r\right\vert }$\\\hline
$\omega^{0}$ & $\left[  \tfrac{\left(  m_{-}-m_{+}\right)  \cos\theta}%
{2}-\tfrac{\left(  R-c\cos\theta\right)  ^{2}+\left(  R\cos\theta-c\right)
\left(  c+m_{-}\right)  }{R-c\cos\theta}\tfrac{k}{R+m_{-}}\right]  d\phi$ &
$\left(  \tfrac{m_{-}-m_{+}}{2}\pm k\right)  d\phi$ ($r\lessgtr R$)\\\hline
$\omega$ & $-\tfrac{e_{-}R\left(  1-\cos\theta\right)  }{2\left(
R+m_{-}\right)  ^{2}}\left[  1+\frac{c^{2}\left(  1+\cos\theta\right)
-R\left(  c+R\allowbreak\right)  }{R\sqrt{R^{2}+c^{2}-2Rc\cos\theta}}\right]
u_{a}u^{a}d\phi$ & $0$\\\hline
$\hat{k}$ & $-\tfrac{e_{-}R\left(  1-\cos\theta\right)  }{2\left(
R+m_{-}\right)  ^{2}}\left[  1+\frac{c^{2}\left(  1+\cos\theta\right)
-R\left(  c+R\allowbreak\right)  }{R\sqrt{R^{2}+c^{2}-2Rc\cos\theta}}\right]
u_{a}u^{a}d\phi$ & $0$\\\hline
$L_{a}$ & $\tfrac{\left(  c+m_{-}\right)  ^{2}}{4km_{-}}l_{a}+u_{a}$ & $u_{a}%
$\\\hline
$L$ & $\tfrac{4ke_{-}^{2}}{\left(  c+m_{-}\right)  ^{3}}\tfrac{R-c}{R+m_{-}%
}u_{a}u^{a}$ & $\tfrac{e_{-}^{2}\left(  R-c\right)  ^{2}}{c^{2}\left(
R+m_{-}\right)  ^{2}}\tfrac{m_{+}\left(  c+m_{-}\right)  ^{2}-4c^{2}k}{\left(
c+m_{-}\right)  ^{3}}u_{a}u^{a}$\\\hline
$\frac{M}{V}$ & $0$ & $0$\\\hline
\end{tabular}
\\
&  \text{Table 5}\colon\text{ Asymptotics at the centers}%
\end{align*}

Note: For some of the functions, the limits towards the Gibbons-Hawking point
are direction dependent; we chose the approach along the positive $z$-axis
($\theta=0$ and $r\rightarrow R$) since that is how the adjacent cycles run.

At infinity we have the limits, to leading orders:%
\begin{align}
V  &  \rightarrow\tfrac{1}{2}\\
\omega^{0}  &  \rightarrow\left[  -\tfrac{\left(  m_{+}-m_{-}\right)
\cos\theta}{2}+k\tfrac{R\cos\theta+m_{-}}{R+m_{-}}\right]  d\phi\\
\omega &  \rightarrow\tfrac{e_{-}}{2}\tfrac{R^{2}-c^{2}}{\left(
R+m_{-}\right)  ^{2}}u_{a}u^{a}\tfrac{\sin^{2}\theta}{r}d\phi\\
\hat{k}  &  \rightarrow\gamma\left[  d\psi-\left(  \tfrac{\left(  m_{+}%
-m_{-}\right)  \cos\theta}{2}-k\tfrac{R\cos\theta+m_{-}}{R+m_{-}}\right)
d\phi\right] \\
L_{a}  &  \rightarrow l_{a}+u_{a}\\
L  &  \rightarrow\tfrac{e_{-}^{2}}{m_{-}^{2}}l_{a}l^{a}+\tfrac{e_{-}^{2}%
}{c^{2}\left(  c+m_{-}\right)  }\left(  \tfrac{m_{+}\left(  R-c\right)
}{R+m_{-}}-\tfrac{4c^{2}k}{\left(  c+m_{-}\right)  ^{2}}\right)  u_{a}u^{a}\\
-\tfrac{M}{V}  &  \rightarrow\tfrac{e_{-}}{m_{-}}l_{a}\left(  l^{a}%
+u^{a}\right)  -\tfrac{e_{-}}{2}\left[  \tfrac{\left(  c+m_{-}\right)
c^{2}-m_{-}m_{+}\left(  R-c\right)  }{c^{2}\left(  R+m_{-}\right)  \left(
c+m_{-}\right)  }+\tfrac{4km_{-}}{\left(  c+m_{-}\right)  ^{3}}\right]
u_{a}u^{a}=\gamma
\end{align}

\subsection{Fields and fluxes}

The fields and fluxes yield at infinity:%
\begin{align}
F^{1}  &  \rightarrow\left(  \tfrac{m_{-}^{2}-c^{2}}{2e_{-}}-\tfrac
{m_{-}\gamma}{\hat{L}}\right)  d\Omega_{2}\\
F^{a}  &  \rightarrow\left[  e_{-}\left(  2l^{a}+u^{a}\right)  -\tfrac
{m_{-}\gamma}{l_{a}+u_{a}}\right]  d\Omega_{2}\\
H_{1}^{\left(  2\right)  }  &  \rightarrow\tfrac{e_{-}}{2}\left(  u_{a}%
\beta^{a}-2\right)  d\Omega_{2}+\left[  \tfrac{1}{4}\left(  2\beta^{a}%
-\tfrac{1}{l_{a}+u_{a}}\right)  \tfrac{\gamma}{l^{a}+u^{a}}-\tfrac{e_{-}%
}{m_{-}}l_{a}\beta^{a}\right]  \left(  d\psi+\omega^{0}\right)
+\text{\textquotedblleft}dt\text{\textquotedblright}\nonumber\\
H_{a}^{\left(  2\right)  }  &  \rightarrow\left(  \tfrac{e_{-}}{2}u_{a}%
\beta^{1}-\tfrac{c^{2}-m_{-}^{2}}{4e_{-}}\eta_{ab}\beta^{b}\right)
d\Omega_{2}+\tfrac{1}{2}\left[  \left(  \tfrac{\gamma}{l^{a}+u^{a}}%
-\tfrac{2e_{-}}{m_{-}}l_{a}\right)  \beta^{1}+\tfrac{\gamma}{\hat{L}}\eta
_{ab}\beta^{b}-\tfrac{\gamma}{\hat{L}\left(  l^{a}+u^{a}\right)  }\right]
\left(  d\psi+\omega^{0}\right)  +\text{\textquotedblleft}%
dt\text{\textquotedblright.}\nonumber
\end{align}

\end{document}